\documentclass{aa}  

\usepackage{graphicx}
\usepackage{txfonts}
\usepackage{graphicx}	
\usepackage{amsmath}	

\usepackage{caption}
\usepackage{subcaption}

\usepackage{booktabs}
\usepackage{threeparttable}

\usepackage[colorlinks=true, linkcolor=blue, citecolor=blue, urlcolor=blue, filecolor=blue,breaklinks=true]{hyperref}

\begin{document}

   \title{REBELS-IFU: Linking damped Lyman-$\alpha$ absorption to [C \textsc{ii}] emission and dust content in the EoR}


   \author{Lucie E. Rowland\inst{1}\fnmsep\thanks{\email{lrowland@strw.leidenuniv.nl}}
  \and Kasper E. Heintz\inst{2,3,4}
  \and Hiddo Algera\inst{5}
  \and Mauro Stefanon\inst{6,7}
  \and Jacqueline Hodge\inst{1}
  \and Rychard Bouwens\inst{1}
  \and Manuel Aravena\inst{8,9}
  \and Elisabete da Cunha\inst{10,11}
  \and Pratika Dayal\inst{12,13,14}
  \and Andrea Ferrara\inst{15}
  \and Rebecca Fisher\inst{16}
  \and Valentino González\inst{17}
  \and Hanae Inami\inst{18}
  \and Olena Komarova\inst{6,7}
  \and Ilse de Looze\inst{22}
  \and Themiya Nanayakkara\inst{19}
  \and Katherine Ormerod\inst{20}
  \and Andrea Pallottini\inst{15,21}
  \and Clara L. Pollock\inst{2, 3}
  \and Renske Smit\inst{20}
  \and Paul van der Werf\inst{1}
  \and Joris Witstok\inst{2,3}
}

\institute{
  Leiden Observatory, Leiden University, P.O. Box 9513, 2300 RA Leiden, The Netherlands 
  \and Cosmic Dawn Center (DAWN), Copenhagen, Denmark 
  \and Niels Bohr Institute, University of Copenhagen, Jagtvej 128, 2200 Copenhagen, Denmark 
  \and Department of Astronomy, University of Geneva, Chemin Pegasi 51, 1290 Versoix, Switzerland 
  \and Institute of Astronomy and Astrophysics, Academia Sinica, 11F of Astronomy-Mathematics Building, No.1, Sec. 4, Roosevelt Rd, Taipei 106319, Taiwan, R.O.C. 
  \and Departament d’Astronomia i Astrofísica, Universitat de València, C. Dr Moliner 50, E-46100 Burjassot, València, Spain 
  \and Unidad Asociada CSIC ‘Grupo de Astrofísica Extragaláctica y Cosmología’ (Instituto de Física de Cantabria – Universitat de València), Spain 
  \and Instituto de Estudios Astrofísicos, Facultad de Ingeniería y Ciencias, Universidad Diego Portales, Av. Ejército 441,  Santiago, Chile 
  \and Millenium Nucleus for Galaxies (MINGAL) 
  \and International Centre for Radio Astronomy Research (ICRAR), The University of Western Australia, Crawley, WA 6009, Australia 
  \and ARC Centre of Excellence for All Sky Astrophysics in 3 Dimensions (ASTRO 3D), Australia 
  \and Canadian Institute for Theoretical Astrophysics, 60 St George St, University of Toronto, Toronto, ON M5S 3H8, Canada 
  \and David A. Dunlap Department of Astronomy and Astrophysics, University of Toronto, 50 St George St, Toronto, ON M5S 3H4, Canada 
  \and Department of Physics, 60 St George St, University of Toronto, Toronto, ON M5S 3H8, Canada 
  \and Scuola Normale Superiore, Piazza dei Cavalieri 7, 56126 Pisa, Italy 
  \and Jodrell Bank Centre for Astrophysics, Department of Physics and Astronomy, University of Manchester, Manchester, UK 
  \and Departamento de Astronomía, Universidad de Chile, Camino del Observatorio 1515, Las Condes, Santiago 7591245, Chile 
  \and Hiroshima Astrophysical Science Center, Hiroshima University, Hiroshima 739-8526, Japan 
  \and Centre for Astrophysics and Supercomputing, Swinburne University of Technology, Hawthorn, VIC 3122, Australia 
  \and Astrophysics Research Institute, Liverpool John Moores University, Liverpool, UK 
  \and Dipartimento di Fisica ``Enrico Fermi'', Università di Pisa, Pisa, Italy 
  \and Sterrenkundig Observatorium, Ghent University, Ghent, Belgium 
}

   \date{Received ; accepted }

 
  \abstract
  {Neutral gas in galaxies during the Epoch of Reionisation (EoR) regulates star formation, dust growth, and the escape of ionising photons, making it a key ingredient in understanding both galaxy assembly and reionisation. Yet, direct constraints on the H\textsc{i} content of galaxies at $z>6$ have been scarce. With \textit{JWST}, Ly$\alpha$ damping wings in galaxy spectra can now provide a direct probe of this neutral component. We analyse \textit{JWST}/NIRSpec prism spectra of 12 UV-luminous galaxies from the REBELS-IFU program at $z\sim6.5$–7.7, deriving H\textsc{i} column densities by modelling Ly$\alpha$ damping wings. Significant damped Ly$\alpha$ absorption is detected in eight galaxies, with $N_{\mathrm{H\textsc{i}}}\gtrsim10^{21}$ cm$^{-2}$. We use the column densities and sizes derived for these sources to estimate their H\textsc{i} mass and compare with $L_{\mathrm{[C \textsc{ii}]}}$–$M_{\mathrm{H\textsc{i}}}$ calibrations. The resulting H\textsc{i} masses show a tentative correlation with those inferred from [C \textsc{ii}], although the [C \textsc{ii}]-based estimates are systematically larger, suggesting that the H\textsc{i} reservoirs may extend beyond the [C \textsc{ii}]-emitting gas. We also combine the DLA-based measurements with FIR-derived dust-to-gas ratios, dust attenuation, and gas-phase metallicities. No correlation is found between DLA-based and FIR-based dust-to-gas ratios, but combining the REBELS-IFU sample with literature samples at lower metallicities reveals a strong correlation between $A_{\mathrm{V}}/N_{\mathrm{H\textsc{i}}}$ and metallicity. These findings suggest that by $z\sim7$ massive galaxies can already host substantial, enriched reservoirs of neutral gas and dust, consistent with $A_{\mathrm{V}}$/$N_{\mathrm{H\textsc{i}}}$–metallicity trends at lower redshift. At the highest redshifts ($z>8$), however, we see tentative evidence for systematically lower $A_{\mathrm{V}}$/$N_{\mathrm{H\textsc{i}}}$ at fixed metallicity, which may point to pristine gas accretion or more efficient dust destruction/expulsion.

  }

    \titlerunning{REBELS-IFU: }
    \authorrunning{Lucie E. Rowland et al.}

   \maketitle
%

\section{Introduction}
\label{sec:intro}

A central goal of galaxy formation theory is to understand how galaxies acquire their gas, form stars, and chemically enrich their interstellar medium (ISM) over cosmic time. This process --  the baryon cycle -- relies critically on the accretion of gas from the intergalactic medium (IGM), its subsequent conversion into molecular gas (the fundamental fuel for star formation), and the return of enriched material through stellar feedback (e.g., \citealt{tumlinson_circumgalactic_2017, tacconi_evolution_2020}). The presence, abundance, and spatial distribution of gas therefore provides vital insight into the physical processes regulating early galaxy growth.

At high redshift, theoretical models and cosmological simulations predict elevated accretion rates of pristine neutral atomic gas onto galaxies, driving an increase in gas turbulence (\citealt{dekel_formation_2009, ginzburg_evolution_2022}), a dilution of gas-phase abundances (e.g., \citealt{ma_origin_2016}), and rapid and bursty star formation (\citealt{pallottini_survey_2022}). Yet, directly measuring the atomic gas content in these early galaxies has remained challenging. The H\textsc{i} 21 cm hyperfine line is typically too faint to be detected for individual galaxies beyond $z \sim 0.5$ with current observational facilities (\citealt{fernandez_highest_2016, sinigaglia_mightee-hi_2022}). Instead, absorption spectroscopy of bright background sources like quasars or gamma-ray bursts (GRBs) has historically been the primary method to study neutral gas at high redshift via damped Lyman-$\alpha$ (Ly$\alpha$) systems (DLAs; e.g., \citealt{wolfe_damped_2005, jakobsson_hi_2006, peroux_cosmic_2020}).

Recently, a new window into H\textsc{i} gas at $z > 6$ has opened, enabled by the James Webb Space Telescope (\textit{JWST}). High-sensitivity rest-UV spectroscopy has revealed broad Ly$\alpha$ absorption in the spectra of star forming galaxies themselves, likely shaped by large columns of neutral gas within or around the galaxies -- extreme DLAs with column densities exceeding $N_{\mathrm{HI}} \gtrsim 10^{22}~\mathrm{cm}^{-2}$ (e.g., \citealt{heintz_extreme_2023,deugenio_jades_2024}, although also see the impact of nebular continuum emission to inferred column densities in \citealt{Katz_21_2024}). These features represent the first direct detections of the neutral atomic gas reservoirs in galaxies during the Epoch of Reionisation (EoR), and challenge earlier assumptions that the damping wings in $z>6$ galaxy spectra solely probe the IGM (\citealt{miralda-escude_reionization_1998, mcquinn_probing_2008}).

These DLAs likely trace the build-up of pristine gas as it accretes from the IGM and fuels early star formation, thus representing a key missing piece in our census of baryonic matter in high-$z$ galaxies. Their presence can also hinder the escape of ionising photons, affecting the contribution of galaxies to reionisation, and bias photometric and spectroscopic redshift measurements based on the Lyman break (e.g., \citealt{heintz_strong_2024, hainline_searching_2024, witstok_witnessing_2025, asada_improving_2025}). Crucially, the shape of the Ly$\alpha$ absorption profile itself can be used to infer the H\textsc{i} column density and constrain the neutral gas content along the line of sight.

While atomic gas at $z>6$ has also been indirectly traced via the [C \textsc{ii}]158$\mu$m far-infrared (FIR) emission line, this line typically probes a mix of ionised, neutral, and molecular ISM phases, and [C \textsc{ii}] luminosity-based conversions into atomic, molecular, or total gas mass are dependent on various ISM properties (e.g., \citealt{vizgan_investigating_2022, casavecchia_atomic_2025, vallini_spatially_2025, khatri_c_2025}). While perhaps more difficult to analyse, the Ly$\alpha$ damping wings therefore offer a complementary and more direct tracer of neutral hydrogen within or around galaxies, especially when combined with other probes of the gas and dust content, key to also understanding where the most prominent far-infrared (FIR) ISM cooling lines like [C \textsc{ii}]158$\mu$m mainly originate from at high redshifts.

In parallel, recent observations have revealed that dust is already widespread by the end of the EoR. ALMA has detected massive dust reservoirs in galaxies out to $z \sim 8.3$ (e.g., \citealt{riechers_dust-obscured_2013,marrone_galaxy_2018, tamura_detection_2019, inami_alma_2022}), suggesting that chemical enrichment and dust build-up occur rapidly after the onset of star formation. However, beyond $z \gtrsim 8$, dust detections become remarkably scarce. This apparent disappearance of dust is particularly intriguing in light of \textit{JWST}’s discovery of an overabundance of UV-bright galaxies at $z > 10$ (\citealt{naidu_two_2022, castellano_early_2023, finkelstein_ceers_2023, harikane_nirspec_2023}), with some studies suggesting that the dust in these galaxies may have been pushed to kpc scales by outflows and/or destroyed by supernovae (\citealt{ferrara_super-early_2024,ferrara_blue_2025}, but see e.g. \citealt{mirocha_balancing_2023, dekel_efficient_2023, trinca_exploring_2024,matteri_can_2025} for alternative explanations). 

A promising route to further understand and directly quantify the build-up of dust and metals in the early Universe is via the dust-to-gas (DTG), dust-to-metal (DTM), and dust-to-stellar (DTS) mass ratios (\citealt{remy-ruyer_gas--dust_2014,vis_using_2017,looze_jingle_2020,galliano_nearby_2021}). These ratios provide key insights into the dominant dust production channels, the efficiency of grain growth in the ISM, and the timescales for chemical enrichment (e.g., see recent review of \citealt{schneider_formation_2024}). By combining new constraints on neutral gas masses with measurements of dust attenuation and metallicity, we can further test theoretical models of early dust production, grain growth, and destruction mechanisms.

In this work, we analyse a sample of massive, UV-luminous star-forming galaxies from the REBELS survey (\citealt{bouwens_reionization_2022}) at $z \sim 6.5-7.7$, with the data described in Section \ref{sec:data}. Using \textit{JWST}/NIRSpec prism spectroscopy, we model their UV continua and Ly$\alpha$ absorption profiles to measure H \textsc{i} column densities and probe the neutral gas content via damped Ly$\alpha$ features (Section \ref{sec:modelling DLA}). We describe the other key ISM properties of this sample in Section \ref{sec:ISM properties}, and derive different estimates for the H\textsc{i} mass in Section \ref{sec:gas masses}. We compare these H\textsc{i} mass estimates to gas masses inferred from [C \textsc{ii}] emission (Section \ref{sec:linking cii and dla}), and explore their relation to dust attenuation and metallicity (Section \ref{sec:DTG}). This multi-tracer approach offers a powerful new window into the early gas and dust reservoirs of galaxies, and sheds a  unique light on how baryons assemble and evolve in the first billion years of cosmic history.

Throughout this work, we assume a standard $\Lambda$CDM cosmology, with $H_0=70$ km s$^{-1}$ Mpc$^{-1}$, $\Omega_m=0.30$ and $\Omega_\Lambda=0.70$. We further adopt a \cite{kroupa_variation_2001} initial mass function, and take solar abundance to be $12+\log(\mathrm{O/H})=8.69$ (\citealt{asplund_chemical_2009}).

\section{Data}
\label{sec:data}

\begin{table*}
\centering
\def\arraystretch{1.25}

\caption{Summary of derived physical properties for the REBELS-IFU sample. The H\,\textsc{i} column densities are derived in this work from DLA fitting to the \textit{JWST}/NIRSpec data. H\,\textsc{i} gas masses are inferred from [C\,\textsc{ii}] luminosities using the calibration derived in \cite{heintz_measuring_2021}. The H\textsc{i} column densities for REBELS-14, 15, 32 and 39 are shown as upper limits due to the low significance of the DLA fit (see Section \ref{sec:modelling DLA} and Appendix \ref{appendix:bad DLA fits}).}

\label{tab:REBELS_properties}
\begin{tabular}{lcccccc}
\toprule
Galaxy & $z_{[\mathrm{CII}]}$ & $A_V$ (mag) & $12+\log(\mathrm{O/H})$ & $\log(L_{[\mathrm{CII}]}/\mathrm{L_\odot})$ & $\log(N_{\mathrm{HI}}/\mathrm{cm}^{-2})$ & $\log(M_{\mathrm{HI,[CII]}}/\mathrm{M_\odot})$  \\
\midrule
REBELS-05 & 6.496 & 0.69$_{-0.04}^{+0.03}$ & 8.51$\pm$0.16 & 8.84$_{-0.06}^{+0.05}$ & 21.24$_{-0.88}^{+0.27}$ & 10.47$\pm$0.19 \\
REBELS-08 & 6.749 & 0.38$_{-0.04}^{+0.03}$ & 8.22$\pm$0.20 & 8.87$_{-0.07}^{+0.06}$ & 21.47$_{-0.36}^{+0.19}$ & 10.76$\pm$0.22 \\
REBELS-12 & 7.349 & 0.45$_{-0.02}^{+0.01}$ & 8.23$\pm$0.13 & 9.01$_{-0.22}^{+0.15}$ & 21.37$_{-0.27}^{+0.17}$ & 10.89$\pm$0.21 \\
REBELS-14 & 7.084 & 0.41$_{-0.02}^{+0.02}$ & 7.90$\pm$0.12 & 8.57$_{-0.11}^{+0.12}$ & $<22.1$ & 10.63$\pm$0.18 \\
REBELS-15 & 6.875 & 0.51$_{-0.02}^{+0.01}$ & 7.78$\pm$0.30 & 8.28$_{-0.11}^{+0.10}$ & $<21.8$ & 10.55$\pm$0.18 \\
REBELS-18 & 7.675 & 0.71$_{-0.02}^{+0.01}$ & 8.50$\pm$0.13 & 9.04$_{-0.03}^{+0.04}$ & 20.94$_{-0.41}^{+0.21}$ & 10.68$\pm$0.15 \\
REBELS-25 & 7.306 & 0.73$_{-0.04}^{+0.04}$ & 8.62$\pm$0.17 & 9.20$_{-0.03}^{+0.03}$ & 22.13$_{-0.46}^{+0.22}$ & 10.74$\pm$0.17 \\
REBELS-29 & 6.685 & 0.45$_{-0.03}^{+0.03}$ & 8.73$\pm$0.15 & 8.74$_{-0.07}^{+0.07}$ & 21.65$_{-0.35}^{+0.19}$ & 10.18$\pm$0.19 \\
REBELS-32 & 6.729 & 0.80$_{-0.05}^{+0.04}$ & 8.48$\pm$0.13 & 8.90$_{-0.05}^{+0.04}$ & $<22.9$ & 10.55$\pm$0.17 \\
REBELS-34 & 6.633 & 0.11$_{-0.03}^{+0.02}$ & 8.33$\pm$0.29 & 8.84$_{-0.14}^{+0.15}$ & 20.99$_{-0.67}^{+0.25}$ & 10.63$\pm$0.17 \\
REBELS-38 & 6.577 & 0.54$_{-0.03}^{+0.03}$ & 8.28$\pm$0.18 & 9.23$_{-0.04}^{+0.04}$ & 21.96$_{-0.20}^{+0.14}$ & 11.07$\pm$0.20 \\
REBELS-39 & 6.847 & 0.24$_{-0.03}^{+0.03}$ & 8.02$\pm$0.29 & 8.90$_{-0.07}^{+0.07}$ & $<22.3$ & 10.97$\pm$0.18 \\
\bottomrule
\end{tabular}

\vspace{5pt}
\begin{tablenotes}
\item{\textbf{Notes:} Col. (1): Galaxy identifier. Col. (2): Spectroscopic redshifts ($z_{\mathrm{[C \textsc{ii}]}}$) from \cite{bouwens_reionization_2022}. Col. (3): Dust attenuation from the SED fitting to the integrated NIRSpec spectra discussed in \cite{rowland_rebels-ifu_2025}, with full details in Stefanon et al. (in prep). Col. (4): Oxygen abundances derived in \cite{rowland_rebels-ifu_2025}. Col. (5): [C \textsc{ii}]158$\mu$m luminosities from Schouws et al. (in prep). Col. (6): H\textsc{i} column densities ($N_{\mathrm{H\textsc{i}}}$) derived in this work. Col. (7): H\textsc{i} gas mass estimates derived in this work from the metallicity-dependent $L_{\mathrm{[C \textsc{ii}]}}$ conversion in \cite{heintz_measuring_2021}. }
\end{tablenotes}
\end{table*}

The 12 galaxies analysed in this work, hereafter referred to as the REBELS-IFU sample, were selected from the Cycle 7 ALMA Large Program (LP) Reionisation-Era Bright Emission Line Survey (REBELS; \citealt{bouwens_reionization_2022}; Schouws et al. in prep), which carried out Band 6 spectral scans targeting the [C \textsc{ii}] 158$\mu$m fine-structure line and underlying dust continuum emission in 36 UV-luminous ($M_{\mathrm{UV,AB}}\lesssim-21.5$ mag) galaxies with photometric redshifts $z_\mathrm{phot} > 6.5$\footnote[1]{A further four sources at $z_{\mathrm{phot}}\gtrsim7.6$ were scanned for [O \textsc{iii}]88$\mu$m emission rather than [C \textsc{ii}], although no lines were detected (\citealt{van_leeuwen_alma_2025}).}. All 12 galaxies within the REBELS-IFU sample have high signal-to-noise (S/N$\gtrsim8$) detections of [C \textsc{ii}], with a spectroscopic redshift range of $6.5 \lesssim z \lesssim 7.7$, and [C \textsc{ii}] luminosities between $\log(L_\mathrm{[C \textsc{ii}]}/L_\odot) = 8.2$--$9.3$ (see Table \ref{tab:REBELS_properties}). All but two of these galaxies (REBELS-15 and REBELS-34) are also detected in the ALMA Band 6 continuum from the LP data (\citealt{inami_alma_2022}).

The REBELS-IFU sample was then observed as part of two Cycle 1 \textit{JWST} NIRSpec/IFU programs: eleven galaxies were observed in GO 1626 (PI M. Stefanon, $\sim 1700$ second exposures per source) and one (REBELS-18) in GO 2659 (PI J. Weaver, on-source exposure of 1.7 hours). These observations used the low-resolution prism mode ($\mathcal{R}\sim100$), covering a $3'' \times 3''$ field of view over an observed wavelength range of $0.6$--$5.3\,\mu$m, enabling detections of key rest-frame optical emission lines, as detailed in \cite{rowland_rebels-ifu_2025}. Full details of the data reduction is presented in Stefanon et al. (in prep), with a summary given in \cite{rowland_rebels-ifu_2025, algera_rebels-ifu_2025, fisher_rebels-ifu_2025}.

\cite{rowland_rebels-ifu_2025} also provide details on the extraction of 1D integrated spectra using masks around each galaxy in the IFU cubes, and the corresponding emission line analysis and metallicity measurements for each source.  In this work, we use the same integrated 1D spectra and, where necessary, the derived emission line fluxes and ISM properties described therein.

\section{Methods}
\label{sec:methods}

\subsection{Modelling the Ly$\alpha$ Damping Wing}
\label{sec:modelling DLA}

To model the observed spectrum of each galaxy, we simultaneously fit the UV continuum (UV slopes, $\beta_{\mathrm{UV}}$, and absolute magnitudes, $M_{\mathrm{UV}}$) simultaneously with absorption by neutral hydrogen. The fit is restricted to rest-frame wavelengths between 1000–2600 \AA, thereby excluding regions affected by strong optical nebular emission lines. We also mask out wavelengths that may be affected by the 2175 \AA~ bump feature, which is detected for some of the REBELS-IFU sources (see \citealt{fisher_rebels-ifu_2025}). We compare the $\beta_{\mathrm{UV}}$ and $M_{\mathrm{UV}}$ values derived simultaneously with the DLA in this work with those derived in \cite{fisher_rebels-ifu_2025}, who mask out $\lambda_{\mathrm{rest}}<1268$\AA~ in Appendix \ref{appendix:comparing UV slopes}.

Following \cite{heintz_extreme_2023}, we account for absorption from the intergalactic medium (IGM) using the formalism of \cite{miralda-escude_reionization_1998} and \cite{totani_implications_2006}, where the IGM transmission redward of Ly$\alpha$ is modelled as:

\begin{align}
\tau_{\mathrm{IGM}}(\lambda_{\mathrm{obs}}, z) &= 
\frac{x_{\mathrm{HI}} R_\alpha \tau_{\mathrm{GP}}(z_{\mathrm{gal}})}{\pi} 
\left( \frac{1 + z_{\mathrm{abs}}}{1 + z_{\mathrm{gal}}} \right)^{3/2} \nonumber \\
&\quad \times \left[ 
I\left( \frac{1 + z_{\mathrm{IGM,u}}}{1 + z_{\mathrm{abs}}} \right) 
- I\left( \frac{1 + z_{\mathrm{IGM,l}}}{1 + z_{\mathrm{abs}}} \right)
\right],
\end{align}

where $I(x)$ is given by Eq. 3 in \cite{totani_implications_2006}, $x_{\mathrm{HI}}$ is the average fraction of neutral hydrogen in the IGM, $z_{\mathrm{gal}}$ is the redshift of the galaxy, $z_{\mathrm{abs}}$ is the redshift of the neutral absorbing gas, and $R_\alpha = \Lambda_\alpha \lambda_\alpha / (4\pi c)$ depends on the Ly$\alpha$ damping constant, $\Lambda_\alpha$, and rest-frame wavelength $\lambda_\alpha = 1216$\,\AA. We set the upper bound at the galaxy redshift, $z_{\mathrm{IGM,u}} = z_{\mathrm{gal}}$, and integrate the expression down to $z_{\mathrm{IGM,u}} = 6$ (as in \citealt{totani_implications_2006, heintz_extreme_2023}). The Gunn–Peterson optical depth is given by Eq. 4 in \cite{totani_implications_2006}, which simplifies to:

\begin{equation}
\tau_{\mathrm{GP}}(z)
\simeq
3.96\times10^5
\left(\frac{1+z}{7}\right)^{3/2},
\end{equation}

for the cosmological parameters assumed in this work.

In addition to IGM absorption, we model the optical depth due to neutral hydrogen along the line-of-sight using a Voigt absorption profile following the analytical approximation derived by \cite{tepper-garcia_voigt_2006},
\begin{equation}
\tau_{\mathrm{ISM}}(\lambda_{\mathrm{obs}}) = C\,a\,H(a,x)\,N_{\ion{H}{i}},
\end{equation}
where $H(a,x)$ is the Voigt-Hjerting function, $a$ is the damping parameter, and $C$ is the photon absorption constant. In this model, the neutral hydrogen column density, $N_{\ion{H}{i}}$, and the absorber redshift, $z_{\mathrm{abs}}$, are the two parameters that describe the DLA. For this work, we assume that the DLA arises from gas in close proximity to the galaxy (e.g. ISM or CGM), and therefore fix the absorber redshift to that of the galaxy ($z_{\mathrm{abs}}=z_{\mathrm{gal}}$), determined from the [C \textsc{ii}] systemic redshifts reported in Table \ref{tab:REBELS_properties}. A key strength of our analysis is that these [C \textsc{ii}] detections from ALMA provide extremely precise spectroscopic redshifts (typically $\Delta z \sim 0.0001$), unlike samples relying solely on \textit{JWST} spectroscopy (often prism), which suffer from larger uncertainties. As discussed in Stefanon et al. (in prep), we also correct for the small wavelength offset between the prism spectra and the [C \textsc{ii}] redshifts, which is a systematic offset also found between prism and grating NIRSpec spectra (e.g. \citealt{deugenio_jades_2025}). 

We also explored leaving $z_{\mathrm{abs}}$ as a free parameter in the fits. Allowing this freedom effectively tests whether the absorption could arise from a foreground system rather than from gas associated with the galaxy itself. In practice, $N_{\mathrm{H\textsc{i}}}$ and $z_{\mathrm{abs}}$ are highly degenerate, particularly at the relatively low spectral resolution and S/N of prism data, and the best-fit absorber redshifts are generally consistent with the systemic [C \textsc{ii}] values within the uncertainties. Only two galaxies (REBELS-12 and REBELS-18) were found to have have both $\Delta z = z_{\mathrm{abs}}-z_{\mathrm{gal}}$ and $N_{\mathrm{HI}}$ values inconsistent with zero (at the $1-1.3\sigma$ level) and prefer a DLA model with $z_{\mathrm{abs}}<z_{\mathrm{gal}}$. It is possible that these objects could lie in overdense regions (REBELS-12 already has one known close neighbour; \citealt{fudamoto_alma_2022}), making foreground absorption plausible (with $\Delta z = 0.35\pm0.20$ and $0.26\pm0.18$, respectively). However, excluding these two galaxies from the analysis does not significantly alter our results: the trends shown in Figures \ref{fig:gas mass comparison} and \ref{fig:DTG_and_metallicity} remain, and in the case of Figure \ref{fig:gas mass comparison}, the correlation even becomes slightly stronger (correlation coefficient 0.68, albeit still not significant with $p$=0.09). Since leaving $z_{\mathrm{abs}}$ free introduces degeneracies and, if interpreted as foreground absorption, would render further steps in our analysis (e.g. linking to [C \textsc{ii}] gas masses or computing DTG) physically inconsistent, we therefore adopt the assumption $z_{\mathrm{abs}}=z_{\mathrm{gal}}$ throughout.

Additionally, we assume $x_{\mathrm{HI}}=0.33$, which was derived in \cite{mason_constraints_2025} for sources at $z\sim5.5-8$, for all 12 of the REBELS-IFU sources. As noted by \cite{huberty_pitfalls_2025}, $x_{\mathrm{HI}}$ and the H\textsc{i} column densities (described below) are degenerate in these fits, which can result in uncertainties of up to 1 dex in low resolution spectra. However, we find that the main conclusions of this work do not significantly change if we instead adopt $x_{\mathrm{HI}}=0.53$, as calculated in \cite{umeda_jwst_2024} for galaxies at $z\sim7$. Quantitatively, adopting  $x_{\mathrm{HI}}=0.53$ results in $N_{\mathrm{H\textsc{i}}}$ values lower by less than 0.1 dex, which is within the uncertainties of the quoted fiducial values in Table \ref{tab:REBELS_properties}. There are also additional degeneracies with Ly$\alpha$ emission, which we do not include in the fitting. Ly$\alpha$ emission has only been detected for three of the sources in the REBELS-IFU sample from MMT/Binospec observations (\citealt{endsley_alma_2022}, discussed below) and is completely blended with the continuum in these prism spectra. Incorporating Ly$\alpha$ emission within the range of Ly$\alpha$ FWHM, velocity offsets and equivalent widths (EWs) obtained from \cite{endsley_alma_2022} also does not significantly impact the derived column densities, however we note the considerable uncertainties due to the unknown contribution of Ly$\alpha$ emission for the majority of this sample. This analysis would therefore benefit from higher resolution rest-UV spectra to further investigate these degeneracies.

Given these broad assumptions, the only parameters that we leave free in our fitting are $N_{\mathrm{H\textsc{i}}}$, $\beta_{\mathrm{UV}}$, and $M_{\mathrm{UV}}$. We fit the full model (UV continuum $\times$ IGM transmission $\times$ ISM damping) to the observed 1D spectrum of each galaxy using the Levenberg–Marquardt minimiser implemented in the \texttt{lmfit} package, minimising the residuals between the model and the observed spectrum. We convolve the full model with the NIRSpec prism spectral resolution ($\mathcal{R}$) prior to fitting, for which we use the empirically-derived resolution curve (Stefanon et al. in prep). At the observed wavelength of Ly$\alpha$ at $z=6.5-7.7$, this results in $\mathcal{R}\sim60$.

\begin{figure}
    \centering
    \includegraphics[width=0.49\textwidth]{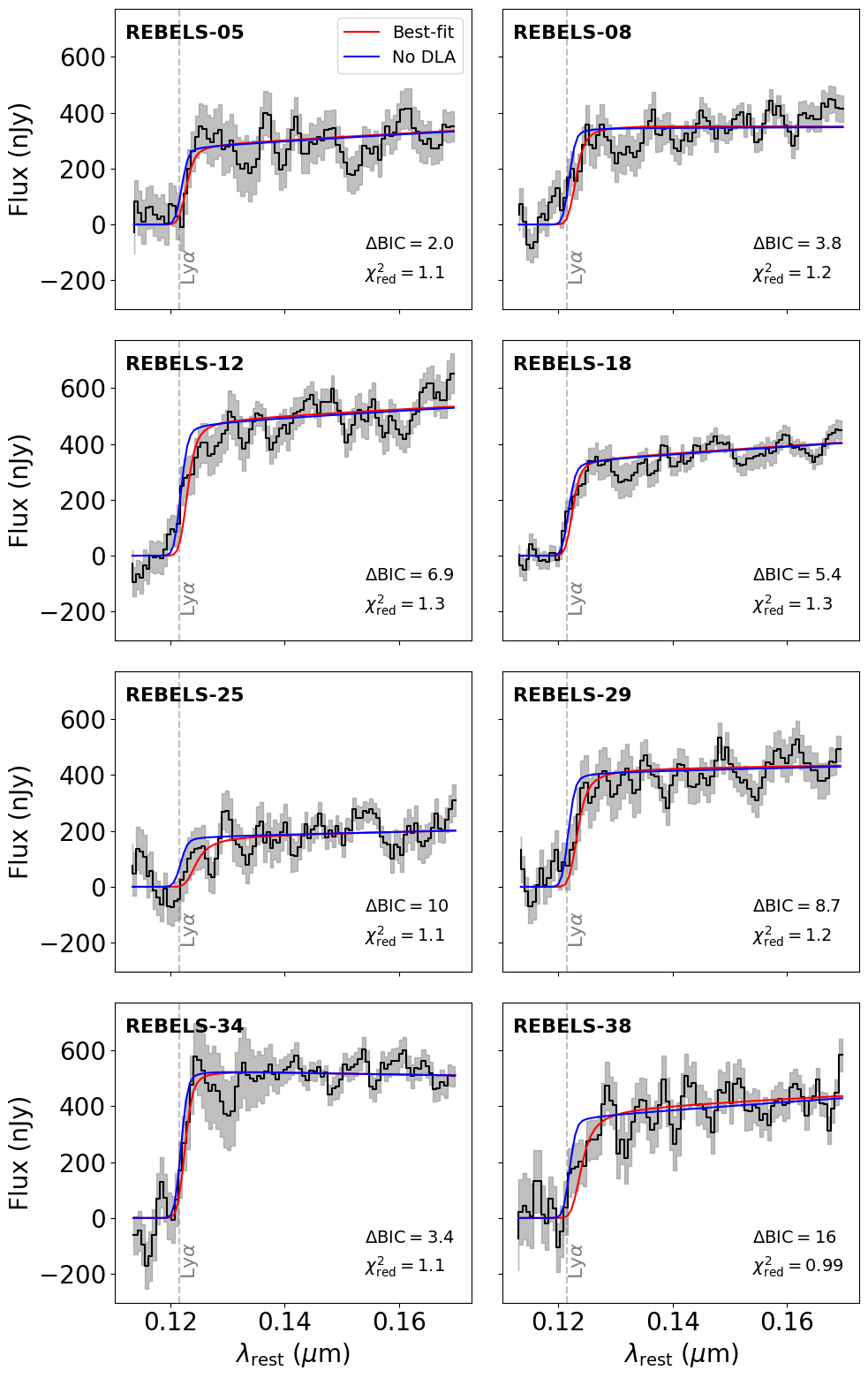}
    \caption{Spectral fits for the eight galaxies where including a damped Ly$\alpha$ (DLA) absorption component improves the fit to the observed UV continuum downturn, as quantified by both the reduced $\chi^2$ and the Bayesian Information Criterion (BIC). The observed data are shown in black, with the flux uncertainties shaded in grey. The best-fit model including both IGM and DLA absorption is plotted in red, while the blue curve shows a model with only IGM absorption.}
    \label{fig:good_dla_fits}
\end{figure}

To assess the significance of the DLA component, we compare the reduced chi-squared ($\chi^2$) and Bayesian Information Criterion (BIC) between the full model and an IGM-only model without the ISM damping component. In the IGM-only case, only the UV continuum slope ($\beta$) and $M_{\mathrm{UV}}$ are fitted, and only absorption from neutral hydrogen in the IGM with a fixed $x_{\mathrm{HI}}=0.33$ is considered. We find that in eight out of 12 sources (plotted in Figure \ref{fig:good_dla_fits}), the addition of the DLA component improves the fit (improved $\chi^2$ and BIC), supporting the presence of high column density neutral hydrogen in the near vicinity of these galaxies. However, we caution that the statistical improvement is only moderate for the majority of sources, as indicated on Figure \ref{fig:good_dla_fits}.

For the remaining four galaxies plotted in Figure \ref{fig:bad_dla_fits} (REBELS-14, 15, 32, and 39), the fits with the DLA component show no statistical improvement. It is interesting to note that three of these galaxies (REBELS-14, 15 and 39) are the only confirmed Ly$\alpha$ emitters in the REBELS-IFU sample (\citealt{endsley_alma_2022}), and have the highest [O \textsc{iii}]$\lambda$5007 EWs and lowest metallicities (\citealt{rowland_rebels-ifu_2025}). We note that when we incorporate the observed FWHM, EW, and velocity offsets of the Ly$\alpha$ emission from \cite{endsley_alma_2022}, there is no statistical improvement to the model with and without a DLA component. This may suggest that the assumption of a uniform H\textsc{i} distribution in our DLA modelling is too simplistic. Indeed the Ly$\alpha$ escape fractions of these galaxies appear to match their inferred LyC escape fractions (Komarova et al., in prep), consistent with escape through low-column “channels” with insufficient H~\textsc{i} for Ly$\alpha$ scattering.

Alternatively, another explanation for the poor DLA fits could be that the observed UV downturns are not due to absorption at all, but rather to intrinsic nebular continuum emission. As noted in Section \ref{sec:intro}, recent studies have shown that strong UV continuum downturns can be produced by two-photon nebular emission in galaxies with extremely hot and young stellar populations (ages $\sim10^6$ yr; $T_\ast \gtrsim 80,000$ K), potentially mimicking the appearance of DLA damping wings (\citealt{Katz_21_2024, Cameron_nebular_2024}). Given their high [O \textsc{iii}]$\lambda$5007 equivalent widths ($\gtrsim 1000$ \AA) and low metallicities ($\lesssim 0.2~\mathrm{Z_\odot}$), such conditions may plausibly apply to REBELS-14, 15 and 39. However, \citet{Katz_21_2024} show that nebular continuum emission under these conditions, and for the measured range of $\beta_{\mathrm{UV}}$ for this sample, would typically mimic DLAs with column densities $N_{\mathrm{H \textsc{i}}} > 10^{22}$ cm$^{-2}$, which is higher than what we infer for these sources. However, note that the inferred DLA column is dependent on the assumed intrinsic SED. \cite{Katz_21_2024} also find that such nebular-dominated spectra are expected to exhibit strong Balmer jumps, which are not observed in our data. We therefore conclude that nebular two-photon emission is unlikely to be the dominant cause of the UV downturns observed in most REBELS galaxies. However, for REBELS-14, 15, and 39, we cannot rule out a moderate contribution from this effect, particularly given their more extreme emission line properties.

Another aspect that could complicate this analysis for the majority of the REBELS-IFU sample is their observed clumpy morphology in the \textit{JWST} emission line maps (see Figure 1 in \citealt{rowland_rebels-ifu_2025}), which could suggest that these are neighbouring or merging galaxies. There is also some evidence that Ly$\alpha$ emission is enhanced in mergers (\citealt{witten_deciphering_2024}), which could provide an explanation for the Ly$\alpha$ emission detected for REBELS-14, 15 and 39. To test the effect this clumpy morphology may have on our results and interpretation, we also fit DLA profiles to individual clump spectra, extracted using the \texttt{ASTRODENDRO}\footnote[1]{\url{http://www.dendrograms.org/}} Python package. Tests using these clump masks on the derived ISM properties (including the oxygen abundance) are discussed in \cite{rowland_rebels-ifu_2025}, with full details of this clump selection and analysis detailed in Rowland et al. (in prep). For the clumps with sufficient S/N, we find that the derived $N_{\mathrm{H\textsc{i}}}$ values are consistent within the uncertainties with the integrated values, although they on average tend to show higher hydrogen column densities (by a factor of $\sim 1.8$). Since one of the aims of this work is to compare the H\textsc{i} gas masses inferred from the DLA fitting to those derived from the [C \textsc{ii}] luminosities, for which the corresponding ALMA data does not have sufficient resolution to identify and analyse individual clumps for the majority of the sample, we therefore only use the $N_{\mathrm{H\textsc{i}}}$ values derived from the DLA fitting of the integrated spectra in subsequent analyses.

For the galaxies with poor fits to the integrated spectra (REBELS-14, 15, 32, and 39), the clump-based analysis likewise does not improve the quality of the DLA fits. Since the models fail to reproduce the apparent UV downturns in REBELS-14, 15, and 39 (both in the integrated and clump spectra), and the UV continuum of REBELS-32 is too faint to reliably constrain a damping wing, we treat all four as non-detections. For these cases, we estimate upper limits for $N_{\mathrm{H\textsc{i}}}$ by iteratively increasing the assumed column density and testing how large $N_{\mathrm{H\textsc{i}}}$ must be for the DLA+IGM models to lie more than $3\sigma$ below the IGM-only models. If we apply this same test to the eight sources with successful DLA fits, only three (REBELS-12, REBELS-29, and REBELS-38) exceed the $3\sigma$ threshold. Thus, while the DLA component improves the fits for most galaxies, the low S/N and modest spectral resolution limit our ability to robustly constrain column densities across the full sample. In Appendix \ref{appendix:bad DLA fits}, we further explore the impact of treating the lower-significance cases as upper limits, where we retain the best-fit column densities for the three robust detections (REBELS-12, 29, and 38) and adopt the 3$\sigma$ upper limits for all other sources.

\subsection{ISM properties}
\label{sec:ISM properties}

The key ISM properties relevant to this study are the oxygen abundance ($12+\log\mathrm{(O/H)}$), and the dust attenuation ($A_\mathrm{V}$). The oxygen abundance values are derived from the same integrated spectra in \cite{rowland_rebels-ifu_2025}, with details discussed therein. In summary, strong line calibrations from \cite{sanders_direct_2024} are used to derive  $12+\log\mathrm{(O/H)}$, primarily using R23 ($(\mathrm{[O \textsc{iii}]}\lambda\lambda 4959,5007 + \mathrm{[O \textsc{ii}]}\lambda\lambda3727,29)/ \mathrm{H\beta}$) and R3 ($=\mathrm{[O \textsc{iii}]\lambda 5007/ \mathrm{H\beta}}$) ratios, and using O32 ($=\mathrm{[O \textsc{iii}]\lambda 5007/[O \textsc{ii}]\lambda\lambda3727,29}$) and/or Ne3O2 ($=\mathrm{[Ne \textsc{iii}]\lambda3869/[O \textsc{ii}]\lambda\lambda3727,29}$) ratios to break the degeneracy for these multi-solution calibrations. 
The resulting metallicities are all $\gtrsim 10\% ~\mathrm{Z_{\odot}}$, with a mean $12+\log(\mathrm{O/H})=8.3$ ($\sim 40\% ~\mathrm{Z_{\odot}}$).

The nebular attenuation values derived from the observed H$\alpha$/H$\beta$ Balmer decrements, $A_{\mathrm{V, neb}}$, are also provided in \cite{rowland_rebels-ifu_2025}. Nebular attenuation values typically exceed the stellar attenuation inferred from spectral energy distribution (SED) fitting, which is mainly based on the attenuation of the stellar continuum, by as much as a factor of two (e.g., \citealt{calzetti_dust_1994}). This is also the case for the REBELS-IFU sample (Fisher et al. in prep). Since the SED-derived $A_{\mathrm{V}}$ represents the integrated attenuation of the whole stellar continuum, and not just in emission-line regions, we adopt SED-derived $A_{\mathrm{V}}$ values for all sources in this work (given in Table \ref{tab:REBELS_properties}), but we note that adopting $A_{\mathrm{V, neb}}$ does not significantly change our findings. The SED fitting of the integrated REBELS-IFU spectra is summarised in \cite{rowland_rebels-ifu_2025}, with full details and values provided in Stefanon et al. (in prep). In short, \texttt{BAGPIPES} is used to fit the observed spectrum of each source, assuming a non-parametric star formation history (SFH) with a continuity prior and a \cite{calzetti_dust_2000} attenuation curve. We also repeat the analysis discussed in subsequent sections with the $A_\mathrm{V}$ values derived from the flexible attenuation curve fitting described in \cite{fisher_rebels-ifu_2025}, and find that any differences are not significant enough to affect the trends discussed in this work.

\subsection{Atomic gas masses}
\label{sec:gas masses}

In this work, we focus on estimates of the atomic (H\textsc{i}) gas mass. As noted in Section \ref{sec:intro}, the [C \textsc{ii}]158$\mu$m line is an oft-used tracer of atomic and/or molecular gas reservoirs within galaxies (e.g. \citealt{pallottini_zooming_2017,vallini_molecular_2017,pineda_herschel_2013,zanella_c_2018,lebouteiller_physical_2019,  madden_tracing_2020, vizgan_investigating_2022, casavecchia_atomic_2025}).  Since both atomic and molecular gas fuel star formation, [C \textsc{ii}] can also trace star formation, although whether its luminosity correlates more fundamentally with gas mass or with the star formation rate remains debated (e.g. \citealt{peng_fine-structure_2025}). More broadly, there remains debate over whether atomic or molecular gas dominates the overall gas budget in the early Universe, and consequently what [C \textsc{ii}] effectively traces (e.g., \citealt{obreschkow_understanding_2009, vallini_cii-sfr_2015, tacconi_phibss_2018,ferrara_physical_2019,pallottini_deep_2019,chowdhury_atomic_2022,    vizgan_investigating_2022, casavecchia_atomic_2025}). We therefore proceed by comparing $L_{\mathrm{[C \textsc{ii}]}}$-based methods for estimating the H\textsc{i} masses in our sample with an independent, DLA-based method (see also \citealt{heintz_massive_2024, deugenio_jades_2024}), and we comment further on these caveats in Section \ref{sec:linking cii and dla} and Appendix \ref{appendix:additional gas mass tests}:

\textbf{(i)} The first method estimates H\textsc{i} masses from the column densities ($N_{\mathrm{HI}}$) inferred through the DLA fitting described in Section \ref{sec:modelling DLA}, which represents the integrated H\textsc{i} abundance in the line-of-sight. We assume a uniform spherical distribution of neutral gas within/around each galaxy, and compute the mass as

\begin{equation}
M_{\mathrm{HI,\,DLA}} = 
m_{\mathrm{H}}\, N_{\mathrm{HI}} \cdot\frac{4}{3}\pi (2^{1/3} \times 1.3 r_e)^2, 
\end{equation}

\noindent where $r_e$ is the measured effective radius of each galaxy,  $N_{\mathrm{HI}}$ is the H\textsc{i} column density, and $m_{\mathrm{H}}$ the mass of a hydrogen atom. For this equation, we assume that the effective radius is equivalent to the half-mass radius, such that the total radius $R=2^{1/3}r_e$, and the additional factor of 1.3 stems from a conversion between the projected 2D effective radius and the 3D half-mass radius for a uniform sphere (i.e., assuming $q_0=1$ in the models of \citealt{price_kinematics_2022}).

Ideally, for $r_e$ we would use the effective radius of the HI emission. Since this is not observable, we investigate the effect of assuming that $r_\mathrm{e,H\textsc{i}} = r_\mathrm{e,[C\textsc{ii}]}$ or $r_\mathrm{e,UV}$. For the [C \textsc{ii}] radii, we fit convolved Sérsic models to the higher spatial resolution (beam FWHM 0.7-3 kpc) [C \textsc{ii}] observations of REBELS-05, -08, -18, -25, -29, and -38 (Phillips et al. in prep, see \citealt{rowland_rebels-25_2024} for REBELS-25). These sizes derived from the image plane are broadly consistent with Gaussian fitting in the $uv$-plane (Astles et al. in prep). For the remaining sources without high resolution follow-up [C \textsc{ii}] data (REBELS-12, 14, 15, 32, 34, 39), we fit exponential profiles to the REBELS ALMA LP observations (resolution $\sim 7$ kpc), with the caveat that these sources are barely resolved and the morphologies are heavily beam-dependent. For the UV sizes, we carry out convolved Sérsic modelling on rest-frame UV maps produced from the \textit{JWST}NIRSpec/IFU cubes (see Figure 1 of \citealt{rowland_rebels-25_2024}), assuming the PSF from \texttt{webbpsf}. However, since the rest-frame UV maps for the majority of sources are clumpy/irregular, we compare single, double, and triple component fits to select the best-fitting models and their total effective radii.  On average, we find that the [C \textsc{ii}] sizes are a factor of $\sim2$ times larger than the UV sizes, consistent with the findings of \cite{fudamoto_alma_2022} from a stacking analysis of the REBELS sources and consistent with findings for the ALPINE/CRISTAL galaxies at $z\sim4-6$ (\citealt{ikeda_alma-cristal_2025}). Full details of the multi-wavelength morphological analysis of the REBELS galaxies from both the REBELS-IFU data and Cycle 2 NIRCam observations of 25 [C \textsc{ii}]-detected REBELS galaxies will be presented in future works. For the subsequent discussions of $M_{\mathrm{HI,~ \mathrm{DLA}}}$, we adopt the [C \textsc{ii}]-based radii, since we expect the [C \textsc{ii}] emission to trace a more extended neutral gas component, whereas UV emission predominantly traces the young stellar populations (e.g., \citealt{fudamoto_alma_2022}). This is discussed in more detail in Section \ref{sec:linking cii and dla}.

\textbf{(ii)} The second method uses a metallicity-dependent conversion between [C \textsc{ii}] luminosity and H \textsc{i} gas mass from \cite{heintz_measuring_2021}, given by

\begin{equation}
\log M_{\mathrm{HI, ~[C \textsc{ii}]}} = 
(-0.87 \pm 0.09) \times \log(Z/Z_\odot) 
+ (1.48 \pm 0.12) + \log L_{\mathrm{[CII]}},
\end{equation}

\noindent where $Z/Z_\odot$ is the gas-phase metallicity relative to solar and $M_{\mathrm{HI}}$ and $L_{\mathrm{[CII]}}$ are in units of $M_\odot$ and $L_\odot$, respectively. This relation is calibrated on the direct measurements of the [C \textsc{ii}]-to-H\textsc{i} conversion factor in star-forming galaxies at $2.19\leq z \leq4.99$ using $\gamma$-ray burst afterglows, and most importantly captures the trend of decreasing [C \textsc{ii}] luminosity per unit gas mass at lower metallicities. This empirical relation has also been reproduced in simulations \citep{vizgan_investigating_2022,casavecchia_atomic_2025}, corroborating its applicability. The resulting neutral gas masses, which we denote as $M_{\mathrm{HI, ~[C \textsc{ii}]}}$, are listed in Table \ref{tab:REBELS_properties} based on [C\,{\sc ii}] luminosities taken from \cite{bouwens_reionization_2022} and Schouws et al. (in prep). 

For method \textbf{(ii)}, which is dependent on $L_{\mathrm{[CII]}}$, we note that there is substantial scatter in the literature regarding the conversion factor between $L_{\mathrm{[CII]}}$ and gas mass.  In recent years, there have also been a number of studies on the dependencies of this conversion factor with metallicity, redshift, and other ISM conditions. While the \cite{heintz_measuring_2021} calibration is adopted as our fiducial approach for estimating $M_{\mathrm{HI}}$ from [C \textsc{ii}], we compare this against other metallicity- or redshift-dependent prescriptions based on cosmological simulations (\citealt{vizgan_investigating_2022, casavecchia_atomic_2025, vallini_spatially_2025, khatri_c_2025}) and analytical models (\citealt{ferrara_physical_2019}) in Appendix \ref{appendix:additional gas mass tests}. In Appendix \ref{appendix:additional gas mass tests}, we also make some comparisons with calibrations that adopt a fixed conversion factor between $L_{\mathrm{[CII]}}$ and either the atomic ($M_{\mathrm{H\textsc{i}}}$), molecular ($M_{\mathrm{H2}}$) or total gas mass (such as \citealt{zanella_c_2018}), but such assumptions likely oversimplify the evolving ISM conditions in the early Universe.

To test these H\textsc{i} mass estimates, this work would benefit from dynamical mass measurements for all galaxies in the sample. Currently, only REBELS-25 has such a constraint ($M_{\rm dyn}=1.2^{+1.0}_{-0.6}\times10^{11} \mathrm{M_{\odot}}$; \citealt{rowland_rebels-25_2024}), which, when compared with its latest stellar mass from \textit{JWST} SED fitting ($M_* \sim 2\times10^9 \mathrm{M_\odot}$, \citealt{rowland_rebels-ifu_2025}), implies a total gas reservoir of order $10^{11}\mathrm{M_\odot}$. The H\textsc{i} masses inferred from both $L_{\rm [CII]}$ and DLA fitting therefore lie  within the allowed dynamical budget, although we note that the gas fraction implied from the kinematics and the stellar mass from the integrated SED fitting is higher than expected for its relatively high metallicity (see \citealt{algera_rebels-ifu_2025}).

\subsection{Dust masses}
\label{sec:dust masses}

The dust masses used in this work are derived in \cite{algera_rebels-ifu_2025}, with the methodology described in detail therein. In summary, ten out of the twelve REBELS-IFU sources are detected in ALMA Band 6 continuum emission, probing rest-frame $\sim160\mu$m (\citealt{inami_alma_2022}). For the two non-detections (REBELS-15 and REBELS-34), we adopt $3\sigma$ upper limits based on the $\sigma_{\mathrm{rms}}$ noise of the continuum images. For all single-band detections, dust masses are inferred assuming a modified blackbody with a fixed dust temperature of $T_{\mathrm{dust}}=45\pm15$ K (consistent with \citealt{sommovigo_alma_2022}) and an emissivity index of $\beta_{\mathrm{IR}}=2.0$. 

Two galaxies in the sample have dust continuum detections in multiple ALMA bands, enabling more robust constraints. REBELS-25 has been detected in six ALMA bands, providing well-constrained values for its dust mass, temperature, and emissivity index (\citealt{algera_accurate_2024}). REBELS-38 is detected in Bands 6 and 8, allowing for a dual-band measurement (\citealt{algera_cold_2024}). For these two sources we adopt the dust masses and temperatures measured directly from the multi-band fits, which give lower $T_{\mathrm{dust}}\sim30$–35 K compared to the assumed $45$ K of the rest of the sample. As a result, REBELS-25 and REBELS-38 are inferred to be significantly more dust-rich, with dust masses higher by $\sim0.4$ dex than what would be obtained under the fixed $T_{\mathrm{dust}}$ assumption. Across the sample, the dust masses span $\log(M_{\mathrm{dust}}/M_\odot)\sim7.0$–8.2, with most of the single-band detections falling in a narrower range of $\sim7.0$–7.3. The dominant source of uncertainty arises from the poorly constrained dust temperatures, for which an uncertainty of $\pm15$~K is propagated through for the $M_{\mathrm{dust}}$ measurements. 

Dust masses for the REBELS sample have also been estimated using alternative methods, e.g. as in \cite{ferrara_alma_2022,sommovigo_alma_2022,dayal_alma_2022}. Overall, the scatter between different methods corresponds to an uncertainty of $\sim0.2-0.4$ dex on individual dust masses.

\section{Results and Discussions}
\label{sec:results}

\subsection{Linking [C \textsc{ii}] emission and damped Ly$\alpha$ profiles}
\label{sec:linking cii and dla}

\begin{figure}
    \centering
    \includegraphics[width=0.49\textwidth]{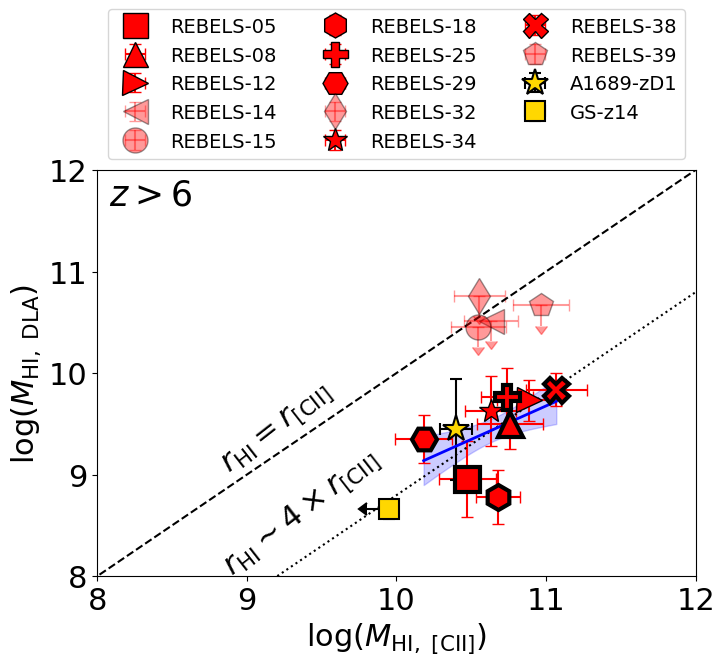}
    \caption{
    Comparison between the H\textsc{i} gas mass inferred from damped Ly$\alpha$ absorption wings ($M_{\mathrm{H\textsc{i},~DLA}}$) and from [C\,\textsc{ii}] luminosities ($M_{\mathrm{H\textsc{i},~[CII]}}$) for the REBELS-IFU sample. Each REBELS galaxy is shown with an individual marker. For REBELS-14, 15, 32, and 39, we use the upper limits derived for the H\textsc{i} column densities. We also plot A1689-zD1 at $z=7.13$ from Heintz et al. (in prep) (yellow star) and GS-z14-0 at $z=14.18$ from \cite{heintz_dissecting_2025-1} and \cite{schouws_deep_2025} (yellow square). For GS-z14-0, we assume $r_{\mathrm{e, [C \textsc{ii}]}}=2\times r_{\mathrm{e, UV}}$, and for A1689-zD1 we determine $r_{\mathrm{e, [C \textsc{ii}]}}$ assuming an exponential profile from the disc diameter reported in Heintz et al. (in prep). The black dashed line shows the one-to-one relation, which assumes that the H \textsc{i} gas reservoir and the [C \textsc{ii}] emission have the same radial extent. The best-fit relation is shown with the solid blue line, with the shaded region indicating the $1\sigma$ uncertainty on the fit. The dotted line shows the average offset in log space, corresponding to a geometric mean mass ratio of $M_{\mathrm{H\textsc{i},~[CII]}} \simeq 22 \times M_{\mathrm{H\textsc{i},~DLA}}$. This implies a typical radius ratio of $r_{\mathrm{H\textsc{i}}} \simeq 4.3-4.7 \times r_{\mathrm{[C \textsc{ii}]}}$, depending on whether the ratio is averaged in log space (4.7) or linear space (4.3). Markers outlined in bold use the [C \textsc{ii}] radii derived from Sérsic fitting to high resolution (beam FWHM $\sim 0.7-3$ kpc) data. For the remaining REBELS sources, the low resolution ($\sim7$ kpc) data is used.} 
    
    \label{fig:gas mass comparison}
\end{figure}

For the first time at $z>6$, we are able to directly compare the [C \textsc{ii}] emission, a common tracer of cold neutral and molecular gas in high-$z$ galaxies, with H\textsc{i} column densities derived from damped Ly$\alpha$ wings. The use of $N_{\mathrm{H\textsc{i}}}$ as a tracer of neutral gas within galaxies is already established out to $z\lesssim5$ (e.g., \citealt{tepper-garcia_voigt_2006, peroux_cosmic_2020}). If [C \textsc{ii}] is an effective tracer of HI within a galaxy, and if the DLA features in this $z>6$ sample are indeed predominantly caused by neutral gas within the ISM or CGM of the galaxy itself, we would expect these observables to be correlated, with an additional dependence on the size of the gas reservoir, and likely also on the metallicity or redshift of the source for the $L_{\mathrm{[CII]}}$-to-$M_{\mathrm{gas}}$ conversions. To encapsulate these additional dependencies, we present a comparison between $M_{\mathrm{HI, ~[C \textsc{ii}]}}$ ($\propto L_{\mathrm{[C \textsc{ii}]}} Z^{-\alpha}$, where $\alpha=0.87$ in \citealt{heintz_measuring_2021}) and $M_{\mathrm{HI, ~DLA}}$ ($\propto N_{\mathrm{H \textsc{i}}} r_e^2$) in Figure \ref{fig:gas mass comparison}. In this figure, we show the eight REBELS-IFU sources with a successful DLA fit, and the upper limits inferred for REBELS-14, 15, 32 and 39. We also plot the results obtained from a similar analysis of A1689-zD1 from Heintz et al. (in prep) and results inferred from the [C \textsc{ii}] upper limits of GS-z14-0 (\citealt{schouws_deep_2025}) and its DLA fitting presented in \cite{heintz_dissecting_2025-1}.

Interestingly, a moderate correlation (Pearson correlation coefficient $=0.5$) is found between these two different H\textsc{i} gas mass estimates. However, the relation is not statistically significant ($p-$value $=0.1$). The small sample size and dynamic range therefore limits our ability to draw firm conclusions, but this tentative relation could indeed indicate that the damping wing and [C \textsc{ii}] emission originate from the same neutral gas reservoir within this sample. Consistent with this picture, \cite{Gelli2025} show from SERRA simulations that DLAs with $N_{\mathrm{HI}}\sim10^{21}\mathrm{cm}^{-2}$ can arise in the galaxy’s immediate surroundings, at distances of only a few kpc.

We also find that the $M_{\mathrm{HI, ~DLA}}$ values are on average a factor of $\sim22 \times$ lower than the $M_{\mathrm{HI, ~[C \textsc{ii}]}}$ values. If the [C \textsc{ii}]-based H\textsc{i} masses are not an overestimate (however see the discussion below and e.g. \citealt{heintz_gas_2023, palla_metal_2024}), this would imply that the radii of the H\textsc{i} gas reservoirs would need to be a factor of $\sim 4.3 \times$ larger than the [C \textsc{ii}] radii ($r_{\mathrm{[CII]}}$) derived in this work. Similarly, if we were to instead use the UV radii to derive  $M_{\mathrm{HI, ~DLA}}$, the H\textsc{i} gas reservoir would need to be on average $\sim8.8\times$ more extended than the UV emission. These estimates depend on the assumed geometry, for which we have adopted a very simplistic scenario, as well as on the $L_{\mathrm{[CII]}}$–$M_{\mathrm{HI}}$ calibration used\footnote[1]{As discussed in more detail in Appendix \ref{appendix:additional gas mass tests}, the other $L_{\mathrm{[C \textsc{ii}]}}$-to-$M_{\mathrm{H\textsc{i}}}$ calibrations tested result in smaller $M_{\mathrm{H\textsc{i}, [C \textsc{ii}]}}$ values, sometimes by as much as $\sim0.4$ dex. This would imply slightly less extended H\textsc{i} gas reservoirs relative to the UV and [C \textsc{ii}] sizes than reported here, but in all cases $M_{\mathrm{H\textsc{i}, [C \textsc{ii}]}}>M_{\mathrm{H\textsc{i}, DLA}}$.}, and should therefore be regarded only as rough indications of the relative extent of the H \textsc{i} gas. Taken together, they support a scenario where the typical central UV star-forming regions in these galaxies are embedded within a larger neutral, atomic gas reservoir that also extends beyond the [C \textsc{ii}]-emitting region.

Indeed, several studies have shown that at high redshift, [C \textsc{ii}] emission is typically more extended than the rest-frame UV continuum, with [C \textsc{ii}] sizes exceeding UV sizes by factors of $\gtrsim2$ (e.g. \citealt{carniani_kiloparsec-scale_2018, fudamoto_alma_2022, fujimoto_alpine-alma_2020, ikeda_alma-cristal_2025}). \cite{ikeda_alma-cristal_2025} further report a negative correlation between [C \textsc{ii}] surface density and Ly$\alpha$ equivalent width, and a tentative anti-correlation between $R_{\rm e,[CII]}/R_{\rm e,UV}$ and Ly$\alpha$ equivalent width, together suggesting that [C \textsc{ii}] traces a more extended atomic gas component.

Comparisons with other cold gas tracers at low redshift also show that [C \textsc{ii}] generally occupies an intermediate scale: more extended than CO but more compact than H\textsc{i} (\citealt{de_blok_comparing_2016,peroux_multiphase_2019,szakacs_muse-alma_2021}). In this context, it is plausible that only the central part of the extended H\textsc{i} reservoir has been significantly metal-enriched, such that [C \textsc{ii}] emission arises only from this enriched inner region, while the more extended outer layer remains relatively metal-poor and therefore faint in [C \textsc{ii}]. Under this assumption, deriving H\textsc{i} masses by equating the radial extent of the H\textsc{i} with that of the [C \textsc{ii}] emission will underestimate the total H\textsc{i} mass that is probed by the DLA absorption.

This picture also provides a useful contrast to absorption-selected studies: for example, \cite{wolfe_nature_2004} used C\textsc{ii}$^\ast$ absorption to infer [C\textsc{ii}] emission and derived very low star-formation surface densities ($\Sigma_{\rm SFR}\sim10^{-3}$–$10^{-2}M_\odot\mathrm{yr}^{-1}\mathrm{kpc}^{-2}$). Such values are far below those we infer for the REBELS galaxies ($\Sigma_{\rm SFR}\sim10 M_\odot\mathrm{yr}^{-1}\mathrm{kpc}^{-2}$; Komarova et al., in prep), and the difference can be understood as a consequence of the regions probed. In absorption-selected studies, the background quasar sightlines most often intersect the relatively unobscured outer ISM/CGM of the foreground galaxy, where the [C \textsc{ii}] emission is likely faint, missing the more central, dusty, star-forming ISM. 

\begin{figure*}
    \centering
    \includegraphics[width=\textwidth]{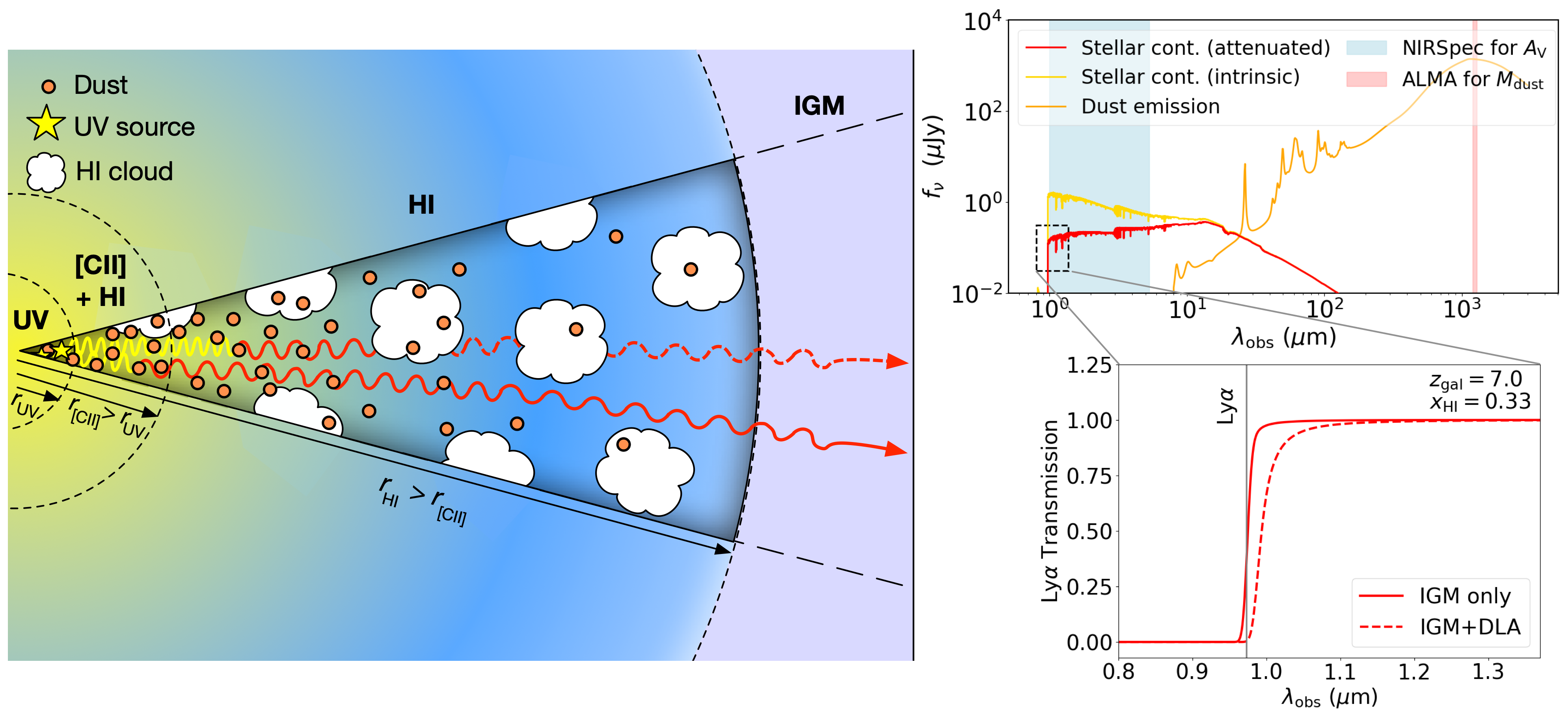}
    \caption{Schematic illustrating the assumptions and methodology used in this work. We stress that this schematic should be regarded only as a rough illustration under the very simplified assumptions adopted. \textit{Left panel:} A simplified model galaxy, where we assume spherical geometry with radii set by the UV emission, [C \textsc{ii}] emission, and the H\textsc{i} reservoir radii derived in this work (see text). Ionising photons from young massive stars (yellow) propagate outward, becoming attenuated by dust (orange circles) and, along some sight-lines, absorbed by neutral H\textsc{i} clouds in the ISM/CGM (dashed red line). Other sightlines pass through without strong absorption (solid red line). All photons are then additionally absorbed by the IGM. \textit{Top right panel:} Example SED from \texttt{BAGPIPES} for a massive ($\log(M_*/\mathrm{M_{\odot}})=9$), dusty ($A_{\mathrm{V}}=0.9$ mag), star-forming ($\mathrm{SFR}_{10}=200 \mathrm{M_{\odot}~yr}^{-1}$) galaxy at $z=7$. Shown are the intrinsic stellar continuum (yellow), the attenuated stellar continuum (red), and dust emission (orange). The blue shaded region marks the NIRSpec wavelength range used to measure stellar $A_V$, while the red shaded band indicates the ALMA Band 6 FIR continuum at $\sim160\mu$m used to estimate the dust mass for the REBELS-IFU sample. \textit{Bottom right panel:} Example Ly$\alpha$ transmission curves, comparing the case of IGM-only absorption (solid red line) with IGM $+$ DLA absorption from neutral gas in the ISM/CGM (dashed red). Following our modelling, we assume a mean IGM neutral fraction of $x_{\mathrm{HI}}=0.33$ and convolve the spectra to $\mathcal{R}=60$.}

    \label{fig:schematic}
\end{figure*}

We illustrate a simplified schematic of these findings in the left panel of Figure \ref{fig:schematic}, where we indicate three concentric spherical regions: a central UV-emitting region, surrounded by a [C \textsc{ii}] emitting region that is $\sim2\times$ more extended, all embedded in a more extended H\textsc{i} gas reservoir. We note that, in this model,  we do not attempt to separate the ISM and CGM. The boundaries between these regions, and the IGM, are not static, and material can mix or be redistributed through inflows, outflows, and environmental interactions. 

Alternatively, it is possible that the \cite{heintz_measuring_2021} may be overestimating $M_{\mathrm{HI, [C\textsc{ii}]}}$. For example, \cite{palla_metal_2024} find that it is not possible to simultaneously reproduce the observed DTS and DTG mass ratios of REBELS galaxies, suggesting a discrepancy in these estimates that could be explained if the gas masses are overestimated. Additionally, \cite{heintz_massive_2024} find that the gas mass of JADES-GS-z14-0 inferred from this calibration is in tension with the inferred dynamical mass, although they caution that the dynamical estimates are uncertain (see also a higher $M_{\mathrm{dyn}}$ estimate from \citealt{scholtz_tentative_2025}). As discussed in Section \ref{sec:gas masses}, we find no such tension between $M_{\mathrm{HI}}$ and $M_{\mathrm{dyn}}$ for REBELS-25, the only REBELS-IFU source with well-resolved kinematics currently available (\citealt{rowland_rebels-25_2024}). Further tests of these mass estimates will be possible with upcoming dynamical constraints for additional REBELS-IFU sources (Phillips et al., in prep.). In Appendix \ref{appendix:additional gas mass tests}, we also derive total, atomic and molecular gas masses using several alternative calibrations. While these yield a wide range of values, across all calibrations we consistently find  $M_{\mathrm{HI, [C\textsc{ii}]}} > M_{\mathrm{HI, DLA}}$, supporting our interpretation that the H\textsc{i} reservoir is more extended than the [C \textsc{ii}]–emitting gas.

\subsection{Dust-to-gas mass ratios}
\label{sec:DTG}

\begin{figure*}
    \centering
    
        \includegraphics[width=\textwidth]{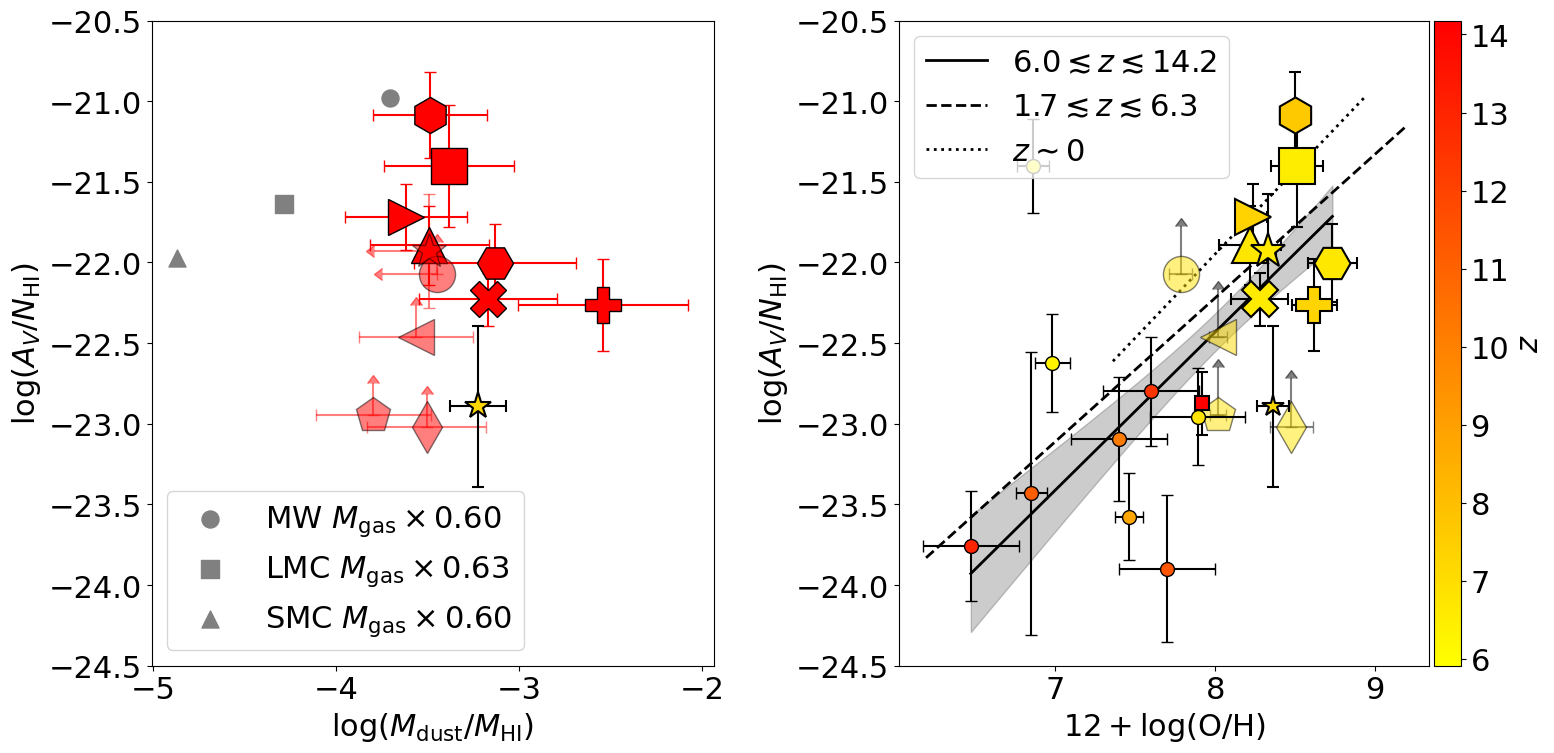}

    \hfill
    
    \caption{
    \textit{Left:} Comparison between the dust-to-gas (DTG) mass ratios derived from ALMA observations (using $L_{\mathrm{[C \textsc{ii}]}}$ to estimate the H\textsc{i} mass and the FIR continuum to infer dust mass) and from DLA-based measurements ($A_{\mathrm{V}} / N_{\mathrm{H\textsc{i}}}$) for the REBELS-IFU galaxies and for A1689-zD1 (Heintz et al. in prep). In both panels, the same marker shapes are used as in the legend of Figure \ref{fig:gas mass comparison}. REBELS-14, 15, 32 and 39 have only upper limits on $N_{\mathrm{H\textsc{i}}}$, and REBELS-34 has an upper limit on the dust mass since it is undetected in Band 6 continuum from the REBELS ALMA LP. No clear trend is observed between these two DTG estimates for the REBELS-IFU sample. We also plot for reference the average values from different sight-lines in the Milky Way (MW), Large Magellanic Cloud (LMC) and Small Magellanic Cloud (SMC) from \cite{konstantopoulou_dust_2024} and references therein, where we adopt their $A_{\mathrm{V,ext}}$ values derived from SED fitting and convert their total gas masses to H\textsc{i} masses using the conversions given in the legend taken from Table 6 of \cite{israel_h_2_1997}. \textit{Right:} Relation between $\log(A_{\mathrm{V}} / N_{\mathrm{H\textsc{i}}})$ and gas-phase metallicity ($12 + \log(\mathrm{O}/\mathrm{H})$) for the REBELS-IFU sample and a compilation of eleven additional $z\gtrsim6$ galaxies from the literature (circle markers), with markers coloured by redshift. The best-fit linear relation is shown via the black solid line, with a shaded region representing the $1\sigma$ uncertainty. A strong correlation is found (Pearson $r = 0.8$, $p = 7\times10^{-5}$), with a consistent slope to the trend reported in \cite{heintz_cosmic_2023} based on GRB sightlines at $z=1.7-6.3$ (black dashed line) and the slope from a linear fit to the data presented in \cite{konstantopoulou_dust_2024} for the Milky Way, LMC and SMC. } 
    
    \label{fig:DTG_and_metallicity}
\end{figure*}

With tentative evidence to support that the damped Ly$\alpha$ wings in this sample of high-$z$ galaxies arise from neutral gas within the galaxies themselves, one might expect a correlation between their dust-to-gas (DTG) mass ratios derived from ALMA observations (using $L_{\mathrm{[ C \textsc{ii}]}}$ to determine the gas mass, and the FIR continuum to trace the dust mass) and those inferred along the line-of-sight via $A_{\mathrm{V}}/ N_{\mathrm{H\textsc{i}}}$. This underlying assumption -- that the same gas reservoir traced by [C \textsc{ii}] and dust traced by FIR emission also produces the DLA absorption and stellar attenuation -- is illustrated schematically in Figure \ref{fig:schematic}.

The left panel of Figure \ref{fig:DTG_and_metallicity} shows this comparison for the eight galaxies in this study with successful DLA fits, as well as one additional source with an upper limit on $N_{\mathrm{H\textsc{i}}}$ and two sources with measurements from the literature (GS-z14-0 from \citealt{heintz_dissecting_2025-1} and A1689-zD1 from Heintz et al. in prep). No clear trend is observed between these two DTG tracers, regardless of whether we adopt method (ii) for converting $L_{\mathrm{[C \textsc{ii}]}}$ to $M_{\mathrm{H\textsc{i}}}$ (Section \ref{sec:gas masses}) or the alternative calibrations discussed in Section \ref{sec:gas masses} and Appendix \ref{appendix:additional gas mass tests}, and regardless of whether we adopt the dust masses from \cite{algera_rebels-ifu_2025} (as plotted in Figure \ref{fig:DTG_and_metallicity} ), \cite{dayal_alma_2022,ferrara_alma_2022} or \citealt{sommovigo_alma_2022}. However, as discussed in Sections \ref{sec:gas masses} and \ref{sec:dust masses}, there are considerable uncertainties in the FIR-based mass measurements. We also note that we only consider the H\textsc{i} mass, without taking into account contributions from molecular gas to the total gas mass. As discussed in Section \ref{sec:gas masses}, the relative contribution of atomic and molecular gas to the total gas budget, and to the observed [C \textsc{ii}] emission, is uncertain at high-$z$.

Although we previously noted tentative evidence for a correlation between $M_{\mathrm{H\textsc{i}}}$ and $N_{\mathrm{H\textsc{i}}}\times R^2$,  we do not find an analogous trend between $M_{\mathrm{dust}}$ and $A_{\mathrm{V}}\times R^2$ (following the assumption in \citealt{carniani_spectroscopic_2024}) for this sample. This outcome is perhaps unsurprising given the well-known limitations of both methods and the small sample sizes considered here. On the FIR side, estimating $M_{\mathrm{dust}}$ from single-band continuum detections carries large uncertainties (see Section \ref{sec:dust masses} and references therein). Similarly, interpreting $A_{\mathrm{V}}$ as a direct tracer of the total dust mass is complicated by unknown dust–star geometry effects. 

The lack of a clear correlation between the FIR-based and DLA-based DTG tracers also mirrors discrepancies reported in studies at lower redshifts. For example, \cite{roman-duval_metal_2022} compared DLA-based DTG ratios (derived from elemental depletion patterns such as [Zn/Fe]) with FIR-based DTG values and found systematic offsets. \cite{konstantopoulou_dust_2024} argue that such discrepancies are likely rooted in the fact that FIR and DLA observations may probe distinct ISM phases, with FIR emission tracing the densest star-forming clouds and DLAs probing more diffuse, extended gas. The finding in Figure \ref{fig:gas mass comparison} discussed in Section \ref{sec:linking cii and dla} that $M_{\mathrm{H\textsc{i}, ~DLA}}< M_{\mathrm{H\textsc{i}, ~[C \textsc{ii}]}}$ may tentatively support this, since this could be interpreted as the H\textsc{i} gas reservoir being more extended than the [C \textsc{ii}] emission. Other DTG studies have also reported discrepancies related to systematic differences between studies using absorption- and emission-line spectroscopy (\citealt{hamanowicz_muse-alma_2020,clark_quest_2023,park_spatially_2024, hamanowicz_metal-z_2024}).

As discussed in Section \ref{sec:intro}, measuring the ratios between dust, gas, stellar, and metal masses across a broad galaxy sample provides key insights into the build-up of dust over cosmic time. At low redshift, strong DTG–metallicity correlations are well established, both in the local Universe (\citealt{remy-ruyer_gas--dust_2014, de_vis_systematic_2019, galliano_nearby_2021}) and at intermediate redshifts $z\simeq1-5$ (\citealt{peroux_cosmic_2020, shapley_first_2020, popping_observed_2022}). At higher redshifts, most theoretical models of early dust enrichment also predict near-linear relations between DTG and metallicity, although the slope and normalisation depend on the prescriptions assumed for supernovae (SNe) dust production, AGB stars, grain growth, and dust destruction (e.g. \citealt{popping_dust_2017, vijayan_detailed_2019, triani_origin_2020,dayal_alma_2022,palla_metal_2024}). Observationally, however, evidence for such correlations at $z>6$ remains weak in current ALMA-based studies, given the limited sample sizes and large uncertainties. In particular, a recent study of the same REBELS-IFU sample (\citealt{algera_rebels-ifu_2025}, notably one of the first $z>6$ DTG and DTM studies), combined with a small number of additional $z\gtrsim6$ galaxies from the literature, found no clear trend between the DTG  ratio and the metallicity of the galaxies when the dust and gas masses were derived from FIR-based detections. However, as discussed therein, this could be driven by the narrow metallicity range probed, the large uncertainties of single-band $M_{\mathrm{dust}}$ estimates, and the limited sample size. Extending observational constraints to a wider range of stellar masses and metallicities is therefore essential for testing these models and advancing our understanding of dust build-up in the early Universe.

If we cautiously proceed with the assumption that $A_{\mathrm{V}}/ N_{\mathrm{H\textsc{i}}}$ can be used as a proxy for the integrated DTG mass ratio, this approach offers a key advantage over FIR tracers: it is more readily accessible in low-mass, metal-poor galaxies -- the typical population at high redshifts. At low stellar masses (and therefore metallicities), detections of [C \textsc{ii}] and the dust continuum become increasingly more challenging. At $z>6$, such FIR detections are generally limited to galaxies with $\log(M_*/\mathrm{M_{\odot}})\gtrsim9$, with the exception of a handful of lensed sources probing lower masses (e.g., \citealt{heintz_gas_2023, fujimoto_primordial_2024}). This limited dynamic range in both stellar mass and metallicity makes it challenging to test theoretical models of dust formation and evolution at early cosmic times. Therefore, by combining the $A_{\mathrm{V}}/ N_{\mathrm{H\textsc{i}}}$ ratios derived in this work with similarly derived ratios from the literature for lower mass sources at $z>6$, we can extend the DTG and metallicity study of \cite{algera_rebels-ifu_2025} over a $\sim 2$ dex range in oxygen abundance, as shown in the right panel of Figure \ref{fig:DTG_and_metallicity}. The literature data points are selected from \cite{watson_dusty_2015, heintz_cosmic_2023, saccardi_dissecting_2023, deugenio_jades_2024, hainline_searching_2024, heintz_dissecting_2025-1, witstok_origins_2025}, and comprise, to the best of our knowledge, all $z>6$ sources with published $N_{\mathrm{HI}}$, $A_{\mathrm{V}}$, and $12+\log(\mathrm{O/H})$ values. 
In these studies, $N_{\mathrm{H\textsc{i}}}$ values are either derived from SED fitting to the full UV spectrum or via a similar DLA fitting as done in this work, and metallicities are derived via nebular emission lines from the galaxy itself (as in this work), or from absorption lines along the line-of-sight.  

A strong correlation emerges between $A_{\mathrm{V}}/ N_{\mathrm{H\textsc{i}}}$ and $12+\log(\mathrm{O/H})$ (with a Pearson correlation coefficient of 0.7 and a $p$-value of $4\times10^{-4}$) when combining the REBELS-IFU sample with the additional eleven sources collected from the literature, which mostly have lower metallicities. 
If we restrict our $A_V/N_{\rm HI}$ comparison to only the REBELS-IFU galaxies (10/13 of the sources in \citealt{algera_rebels-ifu_2025}), our $A_{\mathrm{V}}/N_{\mathrm{H\textsc{i}}}$ analysis yields no statistically significant correlation with metallicity, supporting that the lack of correlation in \cite{algera_rebels-ifu_2025} may indeed be due to the small sample size and lack of dynamic range of the FIR-based measurements.

The best-fit line in the right panel of Figure \ref{fig:DTG_and_metallicity} is given by

\begin{equation}
\log\left(\frac{A_V}{N_{\mathrm{H\textsc{i}}}}\right) = (1.0 \pm 0.2) \times \left[12 + \log(\mathrm{O}/\mathrm{H})\right] - (30 \pm 2).
\end{equation}

\noindent A quantitatively consistent slope is also found in \cite{heintz_cosmic_2023} based on $\gamma$-ray burst sight-lines through star forming galaxies at $z=1.7-6.3$ (dashed black line in the right panel of Figure \ref{fig:DTG_and_metallicity}) and for the Milky Way (MW), Large Magellanic Cloud (LMC), and Small Magellanic Cloud (SMC) in \cite{konstantopoulou_dust_2024} (dotted black line). This consistency in slope across redshift could suggest a common dust production mechanism acting from $z\sim0-8$. For example, models predict an approximately linear relation if SNe dominate dust production (which is expected at high redshift due to the short timescales on which they act, e.g, \citealt{todini_dust_2001}). At higher metallicities ($12+\log{\mathrm{(O/H)}}\gtrsim8$), grain growth in the ISM should steepen the relation (e.g., \citealt{asano_dust_2013, remy-ruyer_gas--dust_2014, zhukovska_dust_2014, galliano_nearby_2021}). This turnover is not clearly seen in the right panel of Figure \ref{fig:DTG_and_metallicity}, nor in the \cite{heintz_cosmic_2023} and \cite{konstantopoulou_dust_2024} samples (although we note limited sampling at the metal-rich end), nor in DLA measurements in \cite{peroux_cosmic_2020} at $z\simeq1-5$. However, as discussed above and in \cite{algera_rebels-ifu_2025}, this discrepancy could be related to systematic differences between studies using absorption- and emission-line spectroscopy (e.g. \citealt{hamanowicz_muse-alma_2020, clark_quest_2023, park_spatially_2024, konstantopoulou_dust_2024, hamanowicz_metal-z_2024}).

Although the slope of the relation appears relatively consistent across redshift, the normalisation differs. For the REBELS-IFU sample at $z\sim6.5$–7.7, we find $A_{\mathrm{V}}/N_{\mathrm{H\textsc{i}}}$ values broadly consistent with those of $z<6$ galaxies at fixed metallicity, suggesting that their ISM is already relatively evolved, consistent with other evidence that the REBELS sample represents a population of evolved, dusty, metal-rich galaxies (\citealt{rowland_rebels-ifu_2025, algera_cold_2024,algera_rebels-ifu_2025, endsley_rebels-mosfire_2025}). However, when considering the full $z\sim6$–14 sample, many high-redshift galaxies (particularly those at $z>8$) lie systematically below the lower-redshift relations (from the aforementioned MW, SMC, LMC and GRB sight-lines), indicating lower $A_{\mathrm{V}}/N_{\mathrm{H\textsc{i}}}$ for their metallicities. This offset may simply reflect sight-line biases: $A_{\mathrm{V}}/N_{\mathrm{H\textsc{i}}}$ is measured along narrow beams and may not capture the global dust–gas distribution. Morphology, dust–star geometry, and metal mixing within the ISM can each bias line-of-sight attenuation estimates. Indeed, spatially resolved studies of nearby galaxies show that DTG can vary by more than an order of magnitude with resolved ISM properties (\citealt{clark_quest_2023,park_spatially_2024}). If the offset is real, however, it could signal intrinsic evolution of the DTG–metallicity relation with redshift. One possibility is that the neutral gas traced by DLAs is systematically more metal-poor than the ionised gas used for emission-line metallicity estimates, perhaps due to pristine IGM accretion (as suggested in \citealt{heintz_dissecting_2025-1}). Another possibility is preferential dust expulsion by feedback or radiation pressure in early galaxies (\citealt{ferrara_blue_2025}), consistent with the low attenuation inferred for many $z>10$ systems.

Ultimately, robust constraints on DTG ratios across cosmic time will require much larger samples spanning a broad mass–metallicity range, with multi-band FIR data and high-S/N UV/optical spectra to robustly constrain $M_{\mathrm{dust}}$, $M_{\mathrm{gas}}$, $A_{\mathrm{V}}$, and $N_{\mathrm{H\textsc{i}}}$. While such datasets remain rare, the REBELS-IFU sample provides an important first step in making these comparisons at $z>6$. Future deep ALMA programs and high-resolution UV spectroscopy will be essential to further these studies.

\section{Summary and Conclusions}
\label{sec:summary}

In this work, we have analysed \textit{JWST}/NIRSpec prism spectra of 12 UV-luminous galaxies from the REBELS-IFU sample at $z\sim6.5-8$ to investigate their neutral gas reservoirs via Ly$\alpha$ damping wing absorption. For eight out of the 12 sources, we find that DLAs with H\textsc{i} column densities $N_{\mathrm{H \textsc{i}}}\gtrsim10^{21}$ cm$^{-2}$ are required to explain the observed UV profiles redwards of Ly$\alpha$. We model these DLAs by accounting for both absorption from the IGM (following \citealt{heintz_extreme_2023, miralda-escude_reionization_1998, totani_implications_2006}) with an average neutral hydrogen fraction in the IGM of $x_{\mathrm{H\textsc{i}}}=0.33$ (\citealt{mason_constraints_2025}), and also absorption from additional neutral hydrogen along the line-of-sight. We fix the redshift of this absorber to the galaxy redshift, assuming that the high column densities inferred arise from H\textsc{i} gas in or around the galaxies themselves (ISM/CGM). 

For the first time, we are able to investigate correlations between H\textsc{i} column densities and [C \textsc{ii}] emission within galaxies in the EoR. By assuming simple spherical geometry, we estimate the mass of neutral gas implied by these column densities within radii equivalent to the extent of the [C \textsc{ii}] emission, and we compare these DLA-derived H\textsc{i} gas masses ($M_{\mathrm{H\textsc{i},DLA}}$) to H\textsc{i} masses derived from the integrated [C \textsc{ii}] luminosities (Figure \ref{fig:gas mass comparison}). 

By combining these two probes of the neutral gas mass with probes of the dust content within these galaxies, we also compare FIR-based DTG ratios with $A_{\mathrm{V}}/N_{\mathrm{H\textsc{i}}}$, which we use to trace the DTG ratio along the line of sight (left panel of Figure \ref{fig:DTG_and_metallicity}). By compiling values of $A_{\mathrm{V}}/N_{\mathrm{H\textsc{i}}}$ at $z>6$ in the literature, we also extend high-$z$ DTG-metallicity relations over a $\sim2$ dex range in oxygen abundance.

The key results of this analysis are:

\begin{itemize}
    \item We find tentative evidence to support that the neutral gas traced by the DLA and the [C \textsc{ii}] emission of massive high-$z$ star-forming galaxies are related, and may be tracing the same neutral gas reservoir within the ISM/CGM of these galaxies. 
    
    \item Whilst there are significant caveats and uncertainties in estimating H\textsc{i} gas masses from both [C \textsc{ii}] emission and $N_{\mathrm{H\textsc{i}}}$, our results suggest that [C \textsc{ii}] emission is more extended than UV emission (by a factor of $\sim2$), but less extended than the H\textsc{i} gas (by a factor of $\sim4$, although this factor is dependent on the $L_{\mathrm{[C \textsc{ii}]}}$-to-$M_{\mathrm{H\textsc{i}}}$ calibration used) in these galaxies. These findings are qualitatively consistent with expectations from lower redshift studies (e.g. \citealt{de_blok_comparing_2016}).
    \item We see no correlation between the FIR emission-based DTG ratios and $A_{\mathrm{V}}/N_{\mathrm{H\textsc{i}}}$ for the REBELS-IFU sample at $z>6$, although we are limited by the small sample size and the uncertainties in the derived properties.
    \item The REBELS-IFU sample exhibits comparable metallicities and $A_{\mathrm{V}}/N_{\mathrm{H\textsc{i}}}$ values with sources at lower redshifts, consistent with other findings that the REBELS-IFU sample represents an evolved sample of massive $z\sim7$ galaxies that already host an enriched ISM with substantial dust and gas reservoirs.
    \item We find a strong correlation between $\log(A_{\mathrm{V}}/N_{\mathrm{H\textsc{i}}})$ and $12+\log(\mathrm{O/H})$ when combining the REBELS-IFU sample with additional, lower metallicity $z>6$ sources from the literature. This near-linear correlation is consistent with expectations from theoretical models of early dust build-up being predominantly driven by SNe (although we cannot rule out significant contribution from efficient ISM grain growth).
    \item We see potential evidence of a redshift evolution in the DTG-metallicity relation, with higher-$z$ sources exhibiting lower $A_{\mathrm{V}}/N_{\mathrm{H\textsc{i}}}$ values for the same metallicity (although see earlier points on potential sight-line biases). If real, this could have a number of interpretations, including that the gas traced by $N_{\mathrm{H\textsc{i}}}$ may contain significant amounts of pristine neutral gas accreted from IGM, and/or more efficient dust destruction/expulsion at high-$z$. Larger samples and more detailed modelling would be necessary to solidify these findings and test these interpretations.
\end{itemize}

Overall, our results highlight that the massive REBELS-IFU galaxies at $z\sim7$ already host substantial and chemically enriched reservoirs of neutral gas and dust, with their UV-bright star-forming regions embedded within more extended atomic gas distributions. This establishes Ly$\alpha$ damping wing fitting as a promising avenue for probing the neutral ISM/CGM in the early Universe, but the present analysis is limited by the modest S/N and spectral resolution of the NIRSpec prism data, as well as by the small sample size. Future progress will require higher-resolution, higher-S/N spectroscopy to better resolve the damping wing profiles, alongside larger samples to test the diversity of neutral gas conditions across galaxy populations. Complementary constraints from resolved [C \textsc{ii}] and other FIR lines, as well as dynamical mass measurements, will be crucial for breaking remaining degeneracies and for building a comprehensive picture of how gas, metals, and dust co-evolve in the first billion years.

\begin{acknowledgements}

LR acknowledges support from the DAWN Visiting Programme, which funded a research visit to the Cosmic Dawn Center (DAWN) that initiated this work. JH acknowledges support from the ERC Consolidator Grant 101088676 (VOYAJ). JAH acknowledges support from the ERC Consolidator Grant 101088676 (VOYAJ). P. Dayal warmly acknowledges support from an NSERC discovery grant (RGPIN-2025-06182). 

This work is based on observations made with the NASA/ESA/CSA \textit{\textit{JWST}}. The data were obtained from the Mikulski Archive for Space Telescopes at the Space Telescope Science Institute, which is operated by the Association of Universities for Research in Astronomy, Inc., under NASA contract NAS 5-03127 for \textit{\textit{JWST}}. These observations are associated with programs \#1626 and \#2659.

This paper made use of the following software packages: Astropy (\citealt{astropy2022}), Matplotlib (\citealt{Hunter:2007}), NumPy (\citealt{harris2020array}), SciPy (\citealt{2020SciPy-NMeth}), spectral-cube (\citealt{spectral-cube}), APLpy (\citealt{aplpy2012}), and astrodendro (http://www.dendrograms.org/).

\end{acknowledgements}


\bibliographystyle{aa} 
\bibliography{REBELS_DLA_v2_bib}

\begin{appendix}

\section{$\beta_{\mathrm{UV}}$ and $M_{\mathrm{UV}}$}
\label{appendix:comparing UV slopes}

\begin{figure*}
    \centering
    \includegraphics[width=0.9\textwidth]{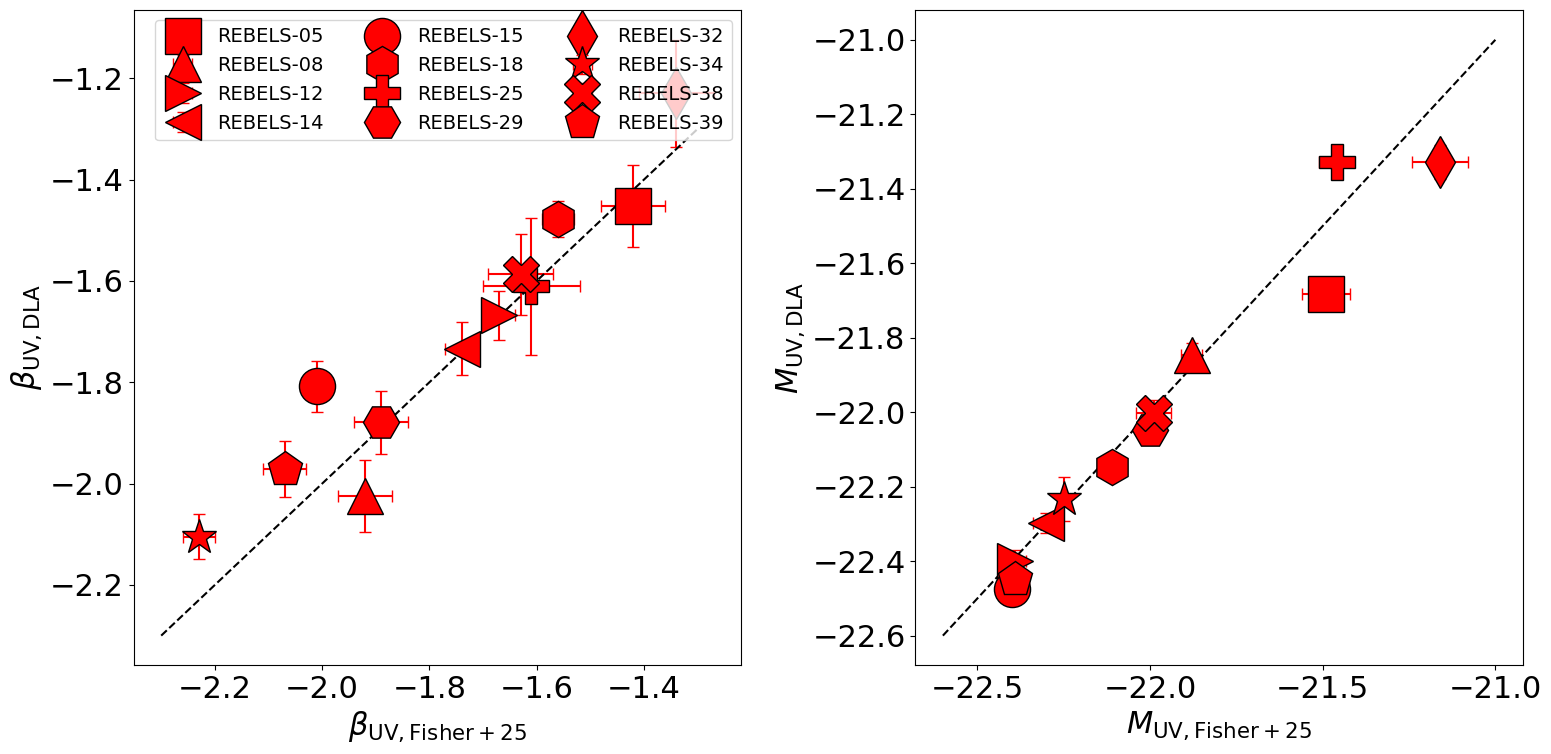}
    \caption{A comparison of the $M_{\mathrm{UV}}$ and $\beta_{\mathrm{UV}}$ values derived from the fitting described in Section \ref{sec:modelling DLA} (where the UV continua are simultaneously modelled with the Ly$\alpha$ absorption) with those derived in \cite{fisher_rebels-ifu_2025}.
    }
    \label{fig:UV_comparison}
\end{figure*}

In Figure \ref{fig:UV_comparison}, we compare $M_{\mathrm{UV}}$ and $\beta_{\mathrm{UV}}$ derived from the fitting described in Section \ref{sec:modelling DLA} (where the UV continua are simultaneously modelled with the Ly$\alpha$ absorption) with those derived in \cite{fisher_rebels-ifu_2025}. There is, in general, good agreement between the two methods. We also note that fixing the $M_{\mathrm{UV}}$ and $\beta_{\mathrm{UV}}$ values to those derived in \cite{fisher_rebels-ifu_2025} does not significantly affect the derived $N_{\mathrm{HI}}$ in the DLA fitting described in Section \ref{sec:modelling DLA}.

\section{Non-significant or poorly constrained DLA fits}
\label{appendix:bad DLA fits}

As discussed in Section \ref{sec:modelling DLA}, only three galaxies in our sample (REBELS-12, 29, and 38) meet our adopted 3$\sigma$ significance threshold, defined as cases where the DLA+IGM model lies more than 3$\sigma$ below the IGM-only model. The remaining sources fall below this threshold, reflecting the limitations imposed by the modest spectral resolution and S/N of the prism data. Four galaxies in particular (REBELS-14, -15, -32, and -39) have column densities consistent with zero and are therefore treated as non-detections. Possible explanations include contributions from two-photon continuum, Ly$\alpha$ emission, merging systems, absorption from a foreground source, or the presence of ionised bubbles around these galaxies. A full exploration of these scenarios will require more detailed modelling and higher-quality data, which is beyond the scope of this work. 

To test the impact of treating the lower-significance cases as non-detections, we repeated our analysis adopting the 3$\sigma$ upper limits for all sources except the three robust detections. The main conclusions remain unchanged. In particular, for the majority of the sources, the upper limits still suggest that $M_{\mathrm{HI, DLA}}<M_{\mathrm{HI, [C \textsc{ii}]}}$ (Figure \ref{fig:gas mass upper limits}) and that the $A_{\mathrm{V}}/N_{\mathrm{HI}}$of the REBELS-IFU sources are still higher than other $z>6$ sources with lower metallicities (Figure \ref{fig:dtg upper limits}).

\begin{figure}
    \centering
    \includegraphics[width=0.49\textwidth]{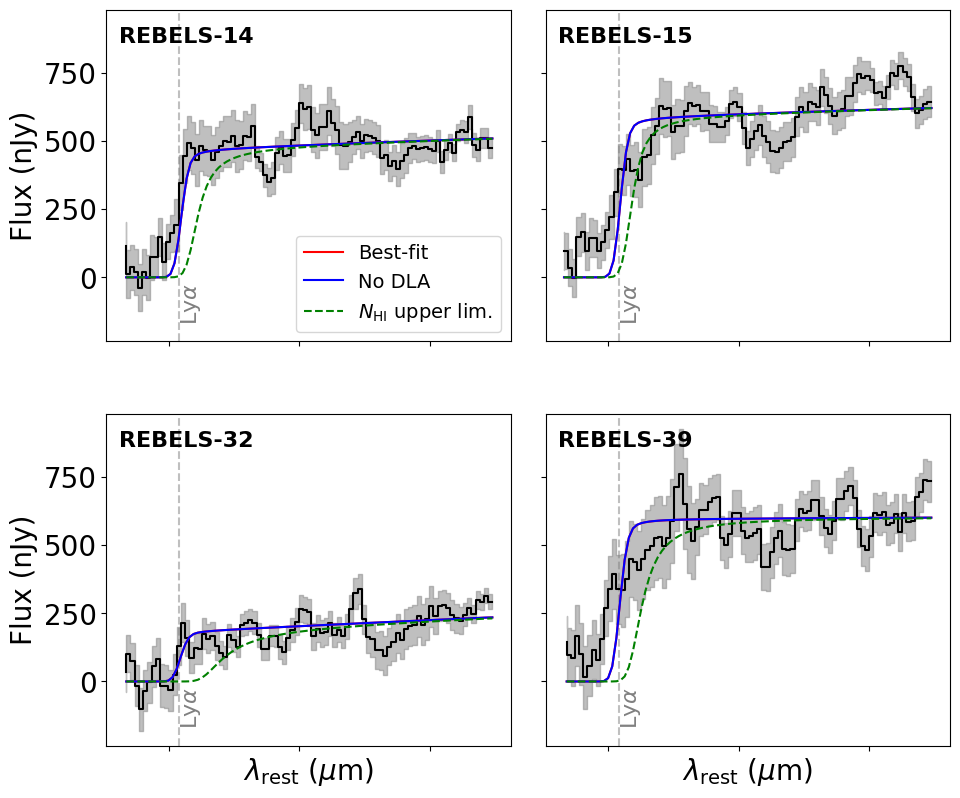}
    \caption{As with Figure \ref{fig:good_dla_fits}, we plot the the best-fit DLA (red) and the IGM-only (blue) model. For these galaxies, the fitted DLA is not a statistical improvement compared to the model with only IGM absorption, and the derived $N_{\mathrm{HI}}$ values are consistent with zero within the uncertainties. We also plot the DLA curve when adopting the upper limit on $N_{\mathrm{NHI}}$ (dashed green curve), which is derived by iteratively increasing $N_{\mathrm{NHI}}$ until the IGM+DLA model lies $3\sigma$ below the IGM-only model at wavelengths close to the Lyman break. 
    }
    \label{fig:bad_dla_fits}
\end{figure}

\begin{figure*}
    \centering
    \begin{subfigure}[b]{0.46\textwidth}
        \includegraphics[width=\textwidth]{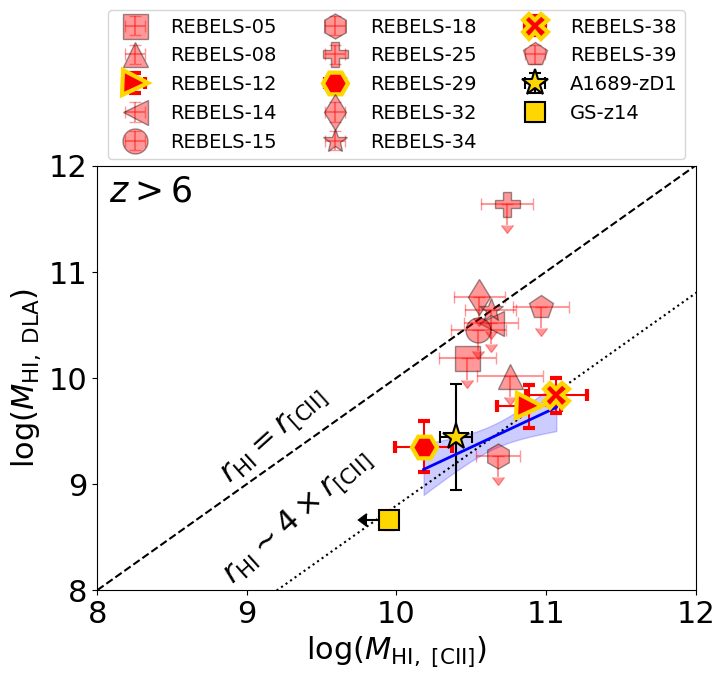}
        
        \caption{}
        \label{fig:gas mass upper limits}
    \end{subfigure}
    \hfill
    \begin{subfigure}[b]{0.46\textwidth}
        \includegraphics[width=\textwidth]{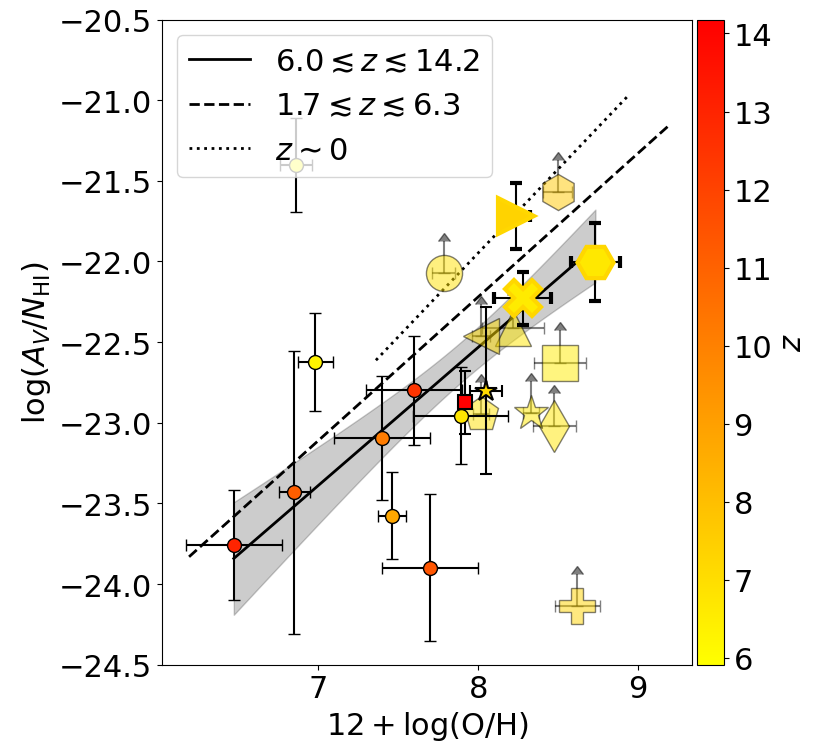}
        
        \caption{}
        \label{fig:dtg upper limits}
    \end{subfigure}
    \caption{\textit{Panel a:} As with Figure \ref{fig:gas mass comparison}, but plotting upper limits for $N_{\mathrm{HI}}$ for all sources apart from those with the most robust DLA fits (REBELS-12, 29, and 38, markers outlined in gold). We also plot the fiducial best-fit line from the text when using the \cite{heintz_measuring_2021} calibrations in blue for comparison. \textit{Panel b:} As with Figure \ref{fig:DTG_and_metallicity}b, but again plotting upper limits for $N_{\mathrm{HI}}$ for all sources apart from those with the most robust DLA fits.}
    \label{fig:upper limits}
\end{figure*}

\section{Additional $L_{\mathrm{[C \textsc{ii}]}}$-to-$M_{\mathrm{gas}}$ calibrations}
\label{appendix:additional gas mass tests}

\begin{figure*}
    \centering
    \begin{subfigure}[b]{0.46\textwidth}
        \includegraphics[width=\textwidth]{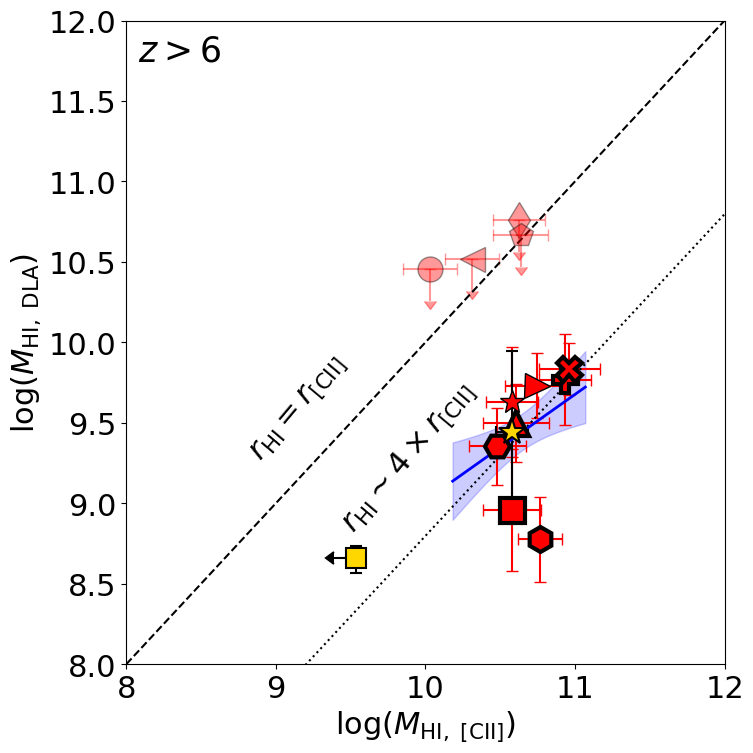}
        
        \caption{}
        \label{fig:vizgan_mhi}
    \end{subfigure}
    \hfill
    \begin{subfigure}[b]{0.46\textwidth}
        \includegraphics[width=\textwidth]{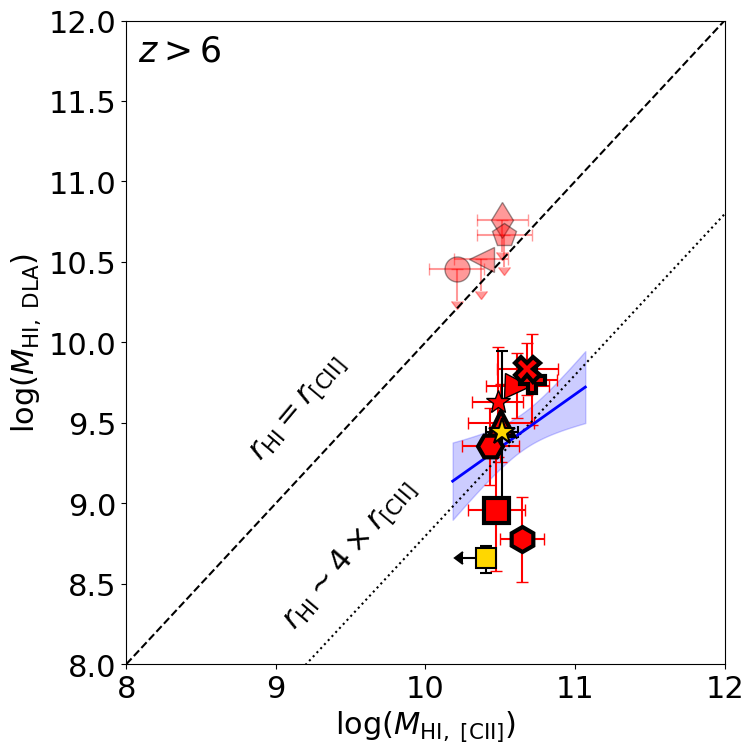}
        
        \caption{}
        \label{fig:casa_mhi}
    \end{subfigure}
    \caption{\textit{Panel a:} As with Figure \ref{fig:gas mass comparison}, but plotting the [C \textsc{ii}]-based H\textsc{i} masses using the calibration from \cite{vizgan_investigating_2022}. We also plot the fiducial best-fit line from the text when using the \cite{heintz_measuring_2021} calibrations in blue for comparison. \textit{Panel b:} The same as the left panel, but using the $L_{\mathrm{[C \textsc{ii}]}}$-to-$M_{\mathrm{H\textsc{i}}}$ calibration from \cite{casavecchia_atomic_2025}.}
    \label{fig:other_mhi_comparisons}
\end{figure*}

\begin{figure*}
    \centering
    \begin{subfigure}[b]{0.46\textwidth}
        \includegraphics[width=\textwidth]{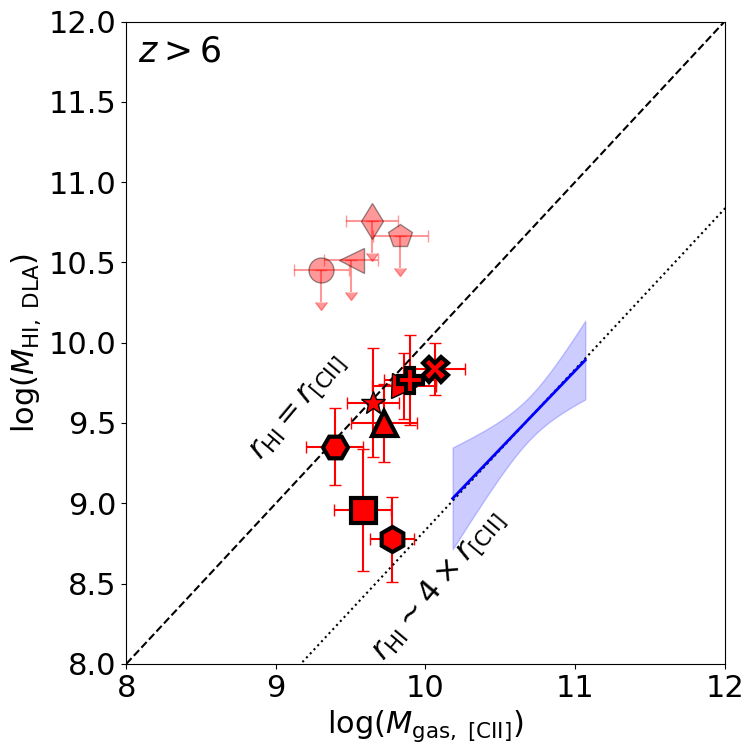}
        
        \caption{}
        \label{fig:vallini_mgas}
    \end{subfigure}
    \hfill
    \begin{subfigure}[b]{0.46\textwidth}
        \includegraphics[width=\textwidth]{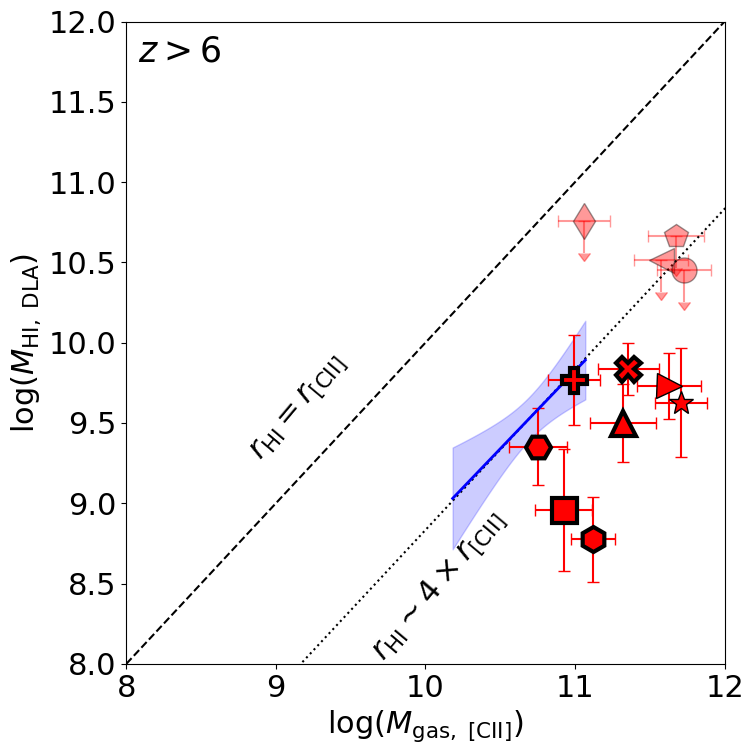}
        
        \caption{}
        \label{fig:ferrara_mgas}
    \end{subfigure}
    \caption{\textit{Panel a:} As with Figures \ref{fig:gas mass comparison} and \ref{fig:other_mhi_comparisons}, but plotting the total gas masses using the $L_{\mathrm{[C \textsc{ii}]}}$-to-$M_{\mathrm{gas}}$ calibration from \cite{vallini_spatially_2025}.  \textit{Panel b:} The same as the left panel, but using the $L_{\mathrm{[C \textsc{ii}]}}$-to-$M_{\mathrm{gas}}$ calibration from \cite{ferrara_physical_2019}.}
    \label{fig:other_mgas_comparisons}
\end{figure*}

In the main text we adopt the metallicity-dependent calibration from \citet{heintz_measuring_2021} to infer H\,\textsc{i} masses from $L_{[\mathrm{CII}]}$. To assess the dependence of the discussion presented in Section \ref{sec:linking cii and dla} on this choice, we repeated the analysis with several alternative prescriptions.

For each galaxy we recomputed gas mass estimates using:
(i) the \cite{vizgan_investigating_2022} relation for $M_{\mathrm{H \textsc{i}}}$, which depends only on $L_{[\mathrm{CII}]}$,
(ii) the \cite{casavecchia_atomic_2025} relation for $M_{\mathrm{H \textsc{i}}}$, which depends on $L_{[\mathrm{CII}]}$ and $z$,
and compared these to our DLA-based H\textsc{i} masses, $M_{\mathrm{H \textsc{i},DLA}}$. We also evaluated four calibrations that do not strictly isolate H\textsc{i}: (iii) \cite{vallini_spatially_2025} (total gas; $M_{\mathrm{gas, ~tot}}= M_{\mathrm{H \textsc{i}}}+M_{\mathrm{H2}} $; depends on $L_{[\mathrm{CII}]}$ and metallicity), (iv) \cite{zanella_c_2018} ($M_{\mathrm{H2}}$ only, with a constant conversion factor of $\alpha_{\mathrm{[C \textsc{ii}]}}=31 \mathrm{M_{\odot}/L_{\odot}}$), (v) \cite{khatri_c_2025} (both $M_{\mathrm{H2}}$ and $M_{\mathrm{gas, ~tot}}$) and (vi) \cite{ferrara_physical_2019} ($M_{\mathrm{gas}}$, dependent on $L_{[\mathrm{CII}]}$, metallicity, and radius). We note that the gas masses derived using the \cite{zanella_c_2018} calibration, which assumes a fixed $\alpha_{\mathrm{[CII]}}=31 \mathrm{M_{\odot} L_{\mathrm{\odot}}^{-1}}$, are presented in \cite{algera_rebels-ifu_2025}. 

Figures \ref{fig:vizgan_mhi} and \ref{fig:casa_mhi} show $M_{\mathrm{HI,DLA}}$ versus $M_{\mathrm{HI,[CII]}}$ for (i) and (ii), respectively, using the same format and markers as Figure \ref{fig:gas mass comparison}. the best-fit relation derived with the \cite{heintz_measuring_2021} calibration (the fiducial values listed in Table \ref{tab:REBELS_properties}) is overplotted in blue in both for reference.

 Averaged over our sample, the alternative $M_{\mathrm{H\textsc{i}}}$ conversions are slightly lower than the values derived using the \cite{heintz_measuring_2021} calibration (by a factor of $\sim0.85$ when using the \cite{vizgan_investigating_2022} calibration and by $\sim 0.69$ when using the \cite{casavecchia_atomic_2025} calibration). However, they are largely consistent within the uncertainties. 

Figures \ref{fig:vallini_mgas} and \ref{fig:ferrara_mgas} show the the values derived using the \cite{vallini_spatially_2025} and the \cite{ferrara_physical_2019} calibrations for the total gas mass, respectively. It would therefore be expected that these are upper limits to $M_{\mathrm{HI}}$. The values derived using the analytical model from \cite{ferrara_physical_2019} (which depends on the $L_{\mathrm{[CII]}}$, metallicity and radius) are on average $\sim 5.5 \times$ higher than the fiducial $M_{\mathrm{HI}}$ values. Conversely, the total gas masses derived from the metallicity-dependent calibration from \cite{vallini_spatially_2025} are on average $\sim0.10\times$ the fiducial H\textsc{i} masses adopted. For REBELS-25, we note that the \cite{vallini_spatially_2025} $M_{\mathrm{gas}}$ estimate is $\sim7.9\times10^{9} \mathrm{M_{\odot}}$, which is $\sim14\times$ lower than the $M_{\mathrm{gas}}$ estimate based on its dynamical and stellar mass ($M_{\mathrm{gas}}=1.1\times10^{11}\mathrm{M_{\odot}}$, \citealt{rowland_rebels-25_2024}).

For completeness, the H$_2$ masses from the \cite{zanella_c_2018} conversion are $\sim0.48\times$ the fiducial H\textsc{i} masses, and the \cite{khatri_c_2025} prescriptions yield $\sim 6 \times$ higher total gas masses and $\sim 30 \times$ higher H$_2$ than $M_{\mathrm{HI}}$ from \cite{heintz_measuring_2021}.

Using the \cite{heintz_measuring_2021} calibration, we find a tentative positive correlation between $M_{\mathrm{HI,[CII]}}$ and $M_{\mathrm{HI,DLA}}$. This correlation weakens for the \cite{vizgan_investigating_2022} and \cite{casavecchia_atomic_2025} prescriptions, both of which do not include an explicit metallicity term. This suggests that incorporating metallicity likely captures physically relevant variations in the [C \textsc{ii}]-to-H\textsc{i} conversion for this sample.

Despite this large range of calibrations, in all cases tested, the [C \textsc{ii}]-based gas mass exceeds the DLA-based H \textsc{i} mass, supporting our interpretation that the H\textsc{i} reservoir is more extended than the region traced by the [C \textsc{ii}] emission.

\end{appendix}

%
%

\end{document}